\documentclass[final,5p,times,twocolumn]{elsarticle}

\usepackage{float}
\usepackage{dblfloatfix}
\usepackage{graphicx} %,subcaption} %[dvips]
\usepackage{tikz}
\usetikzlibrary{positioning,shapes}
\usepackage{pgfplots}
\pgfplotsset{compat=1.15}
\pgfplotsset{minor grid style={gray!30, line width = 0.05}}%dotted}}
\usepackage{bm}
\usepackage{xcolor}
\usepackage{amstext,epsfig,amssymb,amsbsy,amsmath}
\usepackage{mathtools}
\usepackage[output-decimal-marker={.}]{siunitx}
\usepackage{algorithm,algorithmic}
\usepackage{array}
\usepackage{tabularx}
\usepackage{multirow}
\usepackage{multicol}
\usepackage{setspace}
\usepackage{enumitem}
\usepackage{url}
\usepackage[colorlinks=false,allbordercolors=white]{hyperref}
\usepackage{cleveref}
\usepackage{eurosym} % euro sign

\expandafter\def\expandafter\UrlBreaks\expandafter{\UrlBreaks%  save the current one
  \do\a\do\b\do\c\do\d\do\e\do\f\do\g\do\h\do\i\do\j%
  \do\k\do\l\do\m\do\n\do\o\do\p\do\q\do\r\do\s\do\t%
  \do\u\do\v\do\w\do\x\do\y\do\z\do\A\do\B\do\C\do\D%
  \do\E\do\F\do\G\do\H\do\I\do\J\do\K\do\L\do\M\do\N%
  \do\O\do\P\do\Q\do\R\do\S\do\T\do\U\do\V\do\W\do\X%
  \do\Y\do\Z}
  
\newcommand\Tstrut{\rule{0pt}{2.6ex}}         % = `top' strut
\newcommand\Bstrut{\rule[-0.9ex]{0pt}{0pt}}   % = `bottom' strut

\pgfplotsset{
    layers/my layer set/.define layer set={
        background,
        main,
        foreground
    }{},
    set layers=my layer set,
}

\allowdisplaybreaks

\biboptions{numbers,sort&compress}

\journal{Energy Policy}
\begin{document}
\begin{frontmatter}

%% TITLE, AUTHORS AND AFFILIATIONS
\title{\textbf{North~Sea~Energy~Islands: Impact~on~National~Markets~and~Grids}}
\author[inst1]{Andrea~Tosatto}
\author[inst1]{Xavier~Mart\'{i}nez~Beseler}
\author[inst1]{Jacob~{\O}stergaard}
\author[inst3]{Pierre~Pinson}
\author[inst1]{Spyros~Chatzivasileiadis}

\affiliation[inst1]{organization={Department of Electrical Engineering, Technical University of Denmark},
            city={Kongens Lyngby},
            postcode={2800}, 
            country={Denmark}}

% \affiliation[inst2]{organization={Project Management Office, Copenhagen Offshore Partners},
%             city={Copenhagen},
%             postcode={2100}, 
%             country={Denmark}}
            
\affiliation[inst3]{organization={Department of Technology, Management and Economics, Technical University of Denmark},
            city={Kongens Lyngby},
            postcode={2800}, 
            country={Denmark}}

\begin{abstract}
Taking concrete steps towards a carbon-free society, the Danish Parliament has recently approved the establishment of the world's first two offshore energy hubs on Bornholm and on an artificial island in the North Sea. Being the two first-of-their-kind projects, several aspects related to the inclusion of these ``energy islands'' in the current market setup are still under discussion. To this end, this paper presents a first large-scale impact analysis of offshore hubs on the whole European power system and electricity market. Our study shows that energy hubs in the North Sea contribute to increase social welfare in Europe. However, when considering the impact on each country, benefits are not shared equally. To help the development of such projects, we focus on the identification of the challenges arising from the hubs. From a market perspective, we show how exporting countries are affected by the lower electricity prices and we point at heterogeneous consequences induced by new transmission capacity installed in the North Sea. From a system point of view, we show how the large amount of wind energy stresses conventional generators, which are required to become more flexible, and national grids, which cannot always accommodate large imports from the hubs.\\
\\
\noindent
\textit{Corresponding author:} Andrea Tosatto, andrea.tosatto@hotmail.com
\end{abstract}

%% GRAPHICAL ABSTRACT 

%% RESEARCH HIGHLIGHTS
% \begin{highlights}
% \item Energy islands contribute to increase social welfare in Europe
% \item Exporting countries experience a decrease in welfare due to lower electricity prices
% \item New mechanisms have to be introduced to fairly allocate benefits among countries
% \item Unit cycling increases by 11\%, calling for new sources of flexibility
% \item Need for grid reinforcements or storage solutions to absorb the wind energy produced
% \end{highlights}

%% KEYWORDS
\end{frontmatter}

% CHAPTERS %%%%%%%%%%%%%%%%%%%%%%%%%%%%%%%%%%%%%%%%%%%%%%%%%%%%%%%%%%%%%%%%%%%%%%%%%%%%%%%%%%%%%%%%%%%%%%%%%%%%%%%
% \vspace{-0.5em}
\section{Introduction}\label{sec:1}
The Paris Agreement, signed during the Conference of the Parties in 2015 (COP21), represents an important milestone in the pathway toward the transformation of the global energy sector \citep{PARISAGREEMENT}. In order to mitigate climate change and keep global warming 1.5$^\circ$C above pre-industrial levels, ambitious targets related to energy efficiency and Renewable Energy Sources (RES) penetration have been set worldwide. As part of this goal, the European Commission is promoting large investments in solar PV stations and wind farms, which are expected to account for more than 80\% of electricity generation in EU by 2050 \citep{IRENA2020}. 

In this context, large efforts have been made to determine which technologies and geographical areas are the most suitable for accelerating RES penetration \citep{ENEVOLDSEN2019, ELSNER2019}. In the last decades, offshore wind energy has drawn increasing attention, such that it is now considered the most promising technology for achieving RES targets \citep{ESTEBAN2011}. This is mainly due to (i) scarcity of appropriate on-land sites and (ii) public concerns related to noise, visual impact and land utilization \citep{KALDELLIS2013}. Simultaneously, the North Sea region has been identified as the most favorable location for offshore projects because of (i) excellent wind conditions, (ii) shallow waters and (iii) relative proximity to several countries, making it feasible to combine efforts for the realization of large offshore wind installations \citep{SCHILLINGS2012, JONGBLOED2014}. According to \citet{TYNDP2020}, more than 180 GW of wind farms must be installed in the North Sea in order to meet the European targets. 

Together with the development of offshore wind farms, the economic benefits brought by the realization of an offshore transmission grid have been largely analyzed in the literature \citep{SPRO2015, GORENSTEINDEDECCA2016, FLYNN2016, GORENSTEINDEDECCA2018, ALLARD2020}. The first transmission grid project in the North Sea region was proposed by the European Commission in 2008, as the initial step towards the realization of a European super grid. This proposition resulted in the kickoff of the North Seas Countries' Offshore Grid Initiative (NSCOGI) in 2010 \citep{OffshoreGrid}. Ten countries were part of this collaboration: France, Germany, Belgium, the Netherlands, Luxembourg, Denmark, Sweden, Norway, United Kingdom and Ireland. The growing interest around the North Sea region resulted in the integration of NSCOGI into the new North Seas Energy Cooperation (NSEC) in 2016 \citep{NSECdeclaration}. Since then, NSEC has focused on the development of concrete cross-border offshore wind and grid projects (also called ``hybrid'' projects), with the aim of reducing costs and space of offshore developments in the region.

One project that falls within this scope is the ``North Sea Wind Power Hub'' (NSWPH) \citep{NSWPH}. Kicked off in 2017, the NSWPH programme is the result of the joint efforts of Energinet, the Danish electricity and gas Transmission System Operator (TSO), TenneT and Gasunie, respectively the electricity and gas TSOs in the Netherlands and the northern part of Germany. The main innovative aspects of the project are: (i) the concept of ``Hub-and-Spoke'', (ii) the construction of artificial islands instead of traditional offshore platforms and (iii) the modularity of the project. In other words, the project aims at building several small artificial islands (the hubs) where the energy produced by the wind farms is collected and then transmitted to onshore via several High-Voltage Direct-Current (HVDC) links (the spokes).

To facilitate the realization of such hybrid projects, different analyses have been presented in the literature. In \citet{KONSTANTELOS2017}, for example, the authors focus on three specific projects (UK-Benelux, UK-Norway and German Bight) and carry out a detailed cost benefit analysis, proposing methods to allocate benefits among the involved parties. In a similar fashion, the authors in \citet{KITZING2020} present offshore market design options suitable for the future energy islands and study their implications using a simplified model in Balmorel. Moreover, the European Commission has commissioned several studies to engineering consulting firms with the aim of identifying the barriers to the realization of such projects \citep{RolandBerger, THEMA}. In the same vein, the Danish Energy Agency has recently made public two studies related to sea areas screening and cost-benefit analyses of the energy islands \citep{COWI1,COWI2}. All these studies have helped authorities make an informed decision, which led to the Danish Parliament Agreement to construct the first \emph{artificial} energy island in Europe in February 2021 \citep{DANISHNEWS}.

However, most of the studies so far have focused only on the North Sea countries, considering the rest of Europe too far for being impacted by a hub in the North Sea. As pointed out in \citet{RolandBerger}, there is the need for broader analyses to understand the impact of energy islands, as more projects are to be expected. The following questions arise. How are market participants impacted? Is it technically feasible to integrate such a large share of wind generation? The huge amount of installed wind capacity will shift the merit order curve, displacing several conventional generators. In addition, the produced energy will be highly dependent on the weather conditions, and large fluctuations are to be expected. Finally, considering the large amount of transmission capacity planned by the involved TSOs, new market opportunities will arise for generating companies. It is not clear yet how market participants will respond to these changes.

To this end, this paper aims at providing not only the first large-scale impact analysis of the North Sea Energy Islands on the European electricity market and all European countries, but also the complete modelling framework and datasets as an open-access resource. 

More specifically, the contributions of this paper are:
\begin{itemize}
    \item First detailed market simulations studying the impact across all European countries, including the UK, of different combinations of wind and transmission capacity installed in the North Sea.
    \item Evidence and discussion of system needs and market challenges that might arise along with the development of offshore hybrid projects.
    \item The complete modelling framework, including code and datasets, as an open-access resource that can facilitate a wide range of additional studies by researchers and industry professionals on the impact of large-scale projects, including the North Sea Energy Hubs on the European system and electricity markets. The full data are publicly available at: \textcolor{blue}{\url{https://github.com/antosat/European-Transmission-and-Market-Models/}} \citep{GITHUB}.  
\end{itemize}

The studies are carried out using detailed market models of the European electricity market and 400 kV transmission grid in 2030. The market simulations consider the clearing of the day-ahead energy market, formulated as an economic dispatch with grid constraints, where we assume perfect competition and elasticity in the demand. Security constraints for secure system operation are embedded into transmission capacity constraints. The inclusion of energy islands is assessed through the creation of an offshore bidding zone, motivated by the studies in \citet{NSWPHmarket}.

The remainder of this paper is structured as follows. \mbox{Section \ref{sec:2}} discusses market arrangements for hybrid projects. \mbox{Section \ref{sec:3}} presents the modelling assumptions and the market model used for the simulations. In \mbox{Section \ref{sec:4}}, the results of the different simulations are presented and discussed, with particular emphasis on the consequences for market participants. \mbox{Section \ref{sec:5}} gathers conclusions and the policy implications.

% \vspace{-0.5em}
\section{Market arrangements for hybrid projects}\label{sec:2}
\definecolor{yellowSETUP}{rgb}{1,0.752,0}
\definecolor{greenSETUP}{rgb}{0.439,0.678,0.278}
\definecolor{blueSETUP}{rgb}{0.356,0.607,0.835}
\definecolor{orangeSETUP}{rgb}{0.929,0.490,0.192}
\definecolor{redSETUP}{rgb}{0.752,0,0}
\definecolor{darkblueSETUP}{rgb}{0,0.125,0.376}
\begin{figure}[!b]
    \centering
    \resizebox{0.48\textwidth}{!}{%
        \begin{tikzpicture}%[font=\sf]
            % help grid
            % \draw[help lines,step=.2] (0,0) grid (4.5,10);
            % \draw[help lines,line width=.6pt,step=1] (0,0) grid (4.5,10);
            
            %% BOTTOM-LEFT 
            \fill[fill=blueSETUP, draw=none, fill opacity=0.2] (0,0) -- (4.5,0) -- (4.5,0.6) -- (2.25,2.7) -- (0,1.2) -- cycle;
            \fill[fill=greenSETUP, draw=none, fill opacity=0.2] (4.5,0.6) -- (4.5,4.5) -- (2.25,4.5) -- (2.25,2.7) -- cycle;
            \fill[fill=yellowSETUP, draw=none, fill opacity=0.2] (0,1.2) -- (2.25,2.7) -- (2.25,4.5) -- (0,4.5) -- cycle;
            \draw[draw=white, line width = 1] (0,1.2) -- (2.25,2.7) -- (2.25,4.5);
            \draw[draw=white, line width = 1] (4.5,0.6) -- (2.25,2.7);
            \draw[draw=greenSETUP!60, line width=1.5] (4,3.5) -- (2.25,1.9);
            \draw[draw=orangeSETUP, line width=1.5] (0.5,4) -- (2.25,1.9);
            \draw[draw=orangeSETUP, line width=1.5] (2,0.3) -- (2.25,1.9);
            \draw[fill=yellowSETUP, draw=black, line width = 0.5] (0,3) arc[start angle = 270, end angle = 360, radius=1.5] (1.5,4.5) -- (0,4.5) -- (0,3);
            \draw[fill=greenSETUP, draw=black, line width = 0.5] (4.5,2.5) -- (4.5,4.5) -- (3,4.5) arc[start angle = 192, end angle = 241.4, radius=3] -- (4.5,2.5);
            \fill[fill=greenSETUP, draw=black,line width = 0.5, radius=0.4] (2.25,1.9) circle;
            \fill[fill=none, draw=black,line width = 0.5, radius=0.32] (2.25,1.9) circle;
            \draw[fill=blueSETUP, draw=black, line width = 0.5] (0,0) -- (3.5,0) arc[start angle = 0, end angle = 75, radius=0.6] arc[start angle = 75, end angle = 100.2, radius = 7] -- cycle;
            \draw[fill=none, draw=black, line width=0.5] (0,0) rectangle (4.5,4.5);
            \node[anchor=east, fill=greenSETUP!70, draw=greenSETUP, signal, signal to=west, rotate=42, inner sep=1.2pt] at (3.16,3.05) {\textcolor{darkblueSETUP}{\scriptsize 2\tiny GW}};
            \node[anchor=west, fill=orangeSETUP!90, draw=orangeSETUP, signal, signal to=east, rotate=82, inner sep=1.2pt] at (2.32,0.8) {\textcolor{darkblueSETUP}{\scriptsize 2\tiny GW}};
            \node[anchor=east, fill=orangeSETUP!90, draw=orangeSETUP, signal, signal to=west, rotate=-50, inner sep=1.2pt] at (1.48,2.44) {\textcolor{darkblueSETUP}{\scriptsize 4\tiny GW}};
            % \draw[draw=darkblueSETUP, line width=1.5, <-] (2.8,2.7) -- (3.2,3.08);
            % \draw[draw=darkblueSETUP, line width=1.5, ->] (2.35,0.9) -- (2.43,1.32);
            % \draw[draw=darkblueSETUP, line width=1.5, ->] (1.48,2.44) -- (1.1,2.9);
            % \draw[draw=orangeSETUP, line width=1.5] (0.5,4) -- (2.25,1.9);
            % \draw[draw=orangeSETUP, line width=1.5] (2,0.3) -- (2.25,1.9);

            \node at (2.25,2) {\textbf{\scriptsize 0/6}};
            \node at (2.25,1.8) {\tiny GW};
            \node at (1,0.3) {\textbf{\scriptsize BZ3}};
            \node at (1.8,0.3) {\textbf{\scriptsize 5\euro} };
            \node at (0.5,4) {\textbf{\scriptsize BZ1}};
            \node at (0.5,3.6) {\textbf{\scriptsize 10\euro} };
            \node at (4,4) {\textbf{\scriptsize BZ2}};
            \node at (4,3.6) {\textbf{\scriptsize -5\euro} };
            % \node at (2.77,3.1) {\textcolor{darkblueSETUP}{\scriptsize 2\tiny GW}};
            % \node at (2.75,1) {\textcolor{darkblueSETUP}{\scriptsize 2\tiny GW}};
            % \node at (0.95,2.5) {\textcolor{darkblueSETUP}{\scriptsize 4\tiny GW}};
            
            %% CAPTION
            \node[draw=none, fill=none] at (2.25,5) {\small a) Home Market};
            \node[draw=none, fill=none] at (2.25,-0.5) {\small c) Dynamic Low-Price Market};
            
            %% TOP-LEFT
            \fill[fill=blueSETUP, draw=none, fill opacity=0.2] (0,5.5) -- (4.5,5.5) -- (4.5,6.1) -- (2.25,8.2) -- (0,6.7) -- cycle;
            \fill[fill=greenSETUP, draw=none, fill opacity=0.2] (4.5,6.1) -- (4.5,10) -- (2.25,10) -- (2.25,8.2) -- cycle;
            \fill[fill=yellowSETUP, draw=none, fill opacity=0.2] (0,6.7) -- (2.25,8.2) -- (2.25,10) -- (0,10) -- cycle;
            \draw[draw=white, line width = 1] (0,6.7) -- (2.25,8.2) -- (2.25,10);
            \draw[draw=white, line width = 1] (4.5,6.1) -- (2.25,8.2);
            \draw[draw=orangeSETUP, line width=1.5] (2.45,7.65) -- (1.9,7.65);
            \draw[draw=orangeSETUP, line width=1.5] (2.175,7.2) -- (1.9,7.65);
            \draw[draw=orangeSETUP, line width=1.5] (2.175,7.2) -- (2.45,7.65);
            \draw[draw=greenSETUP!60, line width=1.5] (4,9) -- (2.45,7.65);
            \draw[draw=yellowSETUP!60, line width=1.5] (0.5,9.5) -- (1.9,7.65);
            \draw[draw=blueSETUP!60, line width=1.5] (2,5.8) -- (2.175,7.2);
            \draw[fill=yellowSETUP, draw=black, line width = 0.5] (0,8.5) arc[start angle = 270, end angle = 360, radius=1.5] (1.5,10) -- (0,10) -- (0,8.5);
            \draw[fill=greenSETUP, draw=black, line width = 0.5] (4.5,8) -- (4.5,10) -- (3,10) arc[start angle = 192, end angle = 241.4, radius=3] -- (4.5,8);
            \fill[fill=yellowSETUP, draw=black,line width = 0.5, radius=0.2] (1.90,7.65) circle;
            \fill[fill=none, draw=black,line width = 0.5, radius=0.16] (1.90,7.65) circle;
            \fill[fill=greenSETUP, draw=black,line width = 0.5, radius=0.2] (2.45,7.65) circle;
            \fill[fill=none, draw=black,line width = 0.5, radius=0.16] (2.45,7.65) circle;
            \fill[fill=blueSETUP, draw=black,line width = 0.5, radius=0.2] (2.175,7.2) circle;
            \fill[fill=none, draw=black,line width = 0.5, radius=0.16] (2.175,7.2) circle;
            \draw[fill=blueSETUP, draw=black, line width = 0.5] (0,5.5) -- (3.5,5.5) arc[start angle = 0, end angle = 75, radius=0.6] arc[start angle = 75, end angle = 100.2, radius = 7] -- cycle;
            \draw[fill=none, draw=black, line width=0.5] (0,5.5) rectangle (4.5,10);
            \node[anchor=east, fill=greenSETUP!70, draw=greenSETUP, signal, signal to=west, rotate=42, inner sep=1.2pt] at (3.16,8.60) {\textcolor{darkblueSETUP}{\scriptsize 2\tiny GW}};
            \node[anchor=west, fill=white, draw=blueSETUP, signal, signal to=west, rotate=82, inner sep=1.2pt] at (2.32,6.26) {\textcolor{darkblueSETUP}{\scriptsize 3\tiny GW}};
            \node[anchor=east, fill=yellowSETUP!70, draw=yellowSETUP, signal, signal to=west, rotate=-54, inner sep=1.2pt] at (1.38,7.91) {\textcolor{darkblueSETUP}{\scriptsize 4\tiny GW}};
            
            \node at (2.45,7.65) {\textbf{\tiny 0/1} };
            \node at (2.175,7.2) {\textbf{\tiny 3/3} };
            \node at (2.6,7.3) {\textbf{\tiny GW} };
            \node at (1.90,7.65) {\textbf{\tiny 2/2} };
            \node at (1,5.8) {\textbf{\scriptsize BZ3}};
            \node at (1.8,5.8) {\textbf{\scriptsize 5\euro} };
            \node at (0.5,9.5) {\textbf{\scriptsize BZ1}};
            \node at (0.5,9.1) {\textbf{\scriptsize 10\euro} };
            \node at (4,9.5) {\textbf{\scriptsize BZ2}};
            \node at (4,9.1) {\textbf{\scriptsize -5\euro} };
            
        \end{tikzpicture}
            
        \begin{tikzpicture}[font=\sf]
            % help grid
            % \draw[help lines,step=.2] (0,0) grid (4.5,10);
            % \draw[help lines,line width=.6pt,step=1] (0,0) grid (4.5,10);
            
            %% BOTTOM-RIGHT
            \fill[fill=blueSETUP, draw=none, fill opacity=0.2] (0,0) -- (4.5,0) -- (4.5,0.6) -- (2.25,2.7) -- (0,1.2) -- cycle;
            \fill[fill=greenSETUP, draw=none, fill opacity=0.2] (4.5,0.6) -- (4.5,4.5) -- (2.25,4.5) -- (2.25,2.7) -- cycle;
            \fill[fill=yellowSETUP, draw=none, fill opacity=0.2] (0,1.2) -- (2.25,2.7) -- (2.25,4.5) -- (0,4.5) -- cycle;
            \draw[draw=white, line width = 1] (0,1.2) -- (2.25,2.7) -- (2.25,4.5);
            \draw[draw=white, line width = 1] (4.5,0.6) -- (2.25,2.7);
            \draw[draw=orangeSETUP, line width=1.5] (4,3.5) -- (2.25,1.9);
            \draw[draw=orangeSETUP, line width=1.5] (0.5,4) -- (2.25,1.9);
            \draw[draw=orangeSETUP, line width=1.5] (2,0.3) -- (2.25,1.9);
            \draw[fill=yellowSETUP, draw=black, line width = 0.5] (0,3) arc[start angle = 270, end angle = 360, radius=1.5] (1.5,4.5) -- (0,4.5) -- (0,3);
            \draw[fill=greenSETUP, draw=black, line width = 0.5] (4.5,2.5) -- (4.5,4.5) -- (3,4.5) arc[start angle = 192, end angle = 241.4, radius=3] -- (4.5,2.5);
            \fill[fill=red, draw=black,line width = 0.5, radius=0.4] (2.25,1.9) circle;
            \fill[fill=none, draw=black,line width = 0.5, radius=0.32] (2.25,1.9) circle;
            \draw[fill=blueSETUP, draw=black, line width = 0.5] (0,0) -- (3.5,0) arc[start angle = 0, end angle = 75, radius=0.6] arc[start angle = 75, end angle = 100.2, radius = 7] -- cycle;
            \draw[fill=none, draw=black, line width=0.5] (0,0) rectangle (4.5,4.5);
            \node[anchor=east, fill=orangeSETUP!90, draw=orangeSETUP, signal, signal to=west, rotate=42, inner sep=1.2pt] at (3.16,3.05) {\textcolor{darkblueSETUP}{\scriptsize 2\tiny GW}};
            \node[anchor=west, fill=white, draw=blueSETUP, signal, signal to=west, rotate=82, inner sep=1.2pt] at (2.32,0.76) {\textcolor{darkblueSETUP}{\scriptsize 4\tiny GW}};
            \node[anchor=east, fill=orangeSETUP!90, draw=orangeSETUP, signal, signal to=west, rotate=-50, inner sep=1.2pt] at (1.48,2.44) {\textcolor{darkblueSETUP}{\scriptsize 4\tiny GW}};
            
            \node at (2.25,2) {\textbf{\scriptsize 6/6}};
            \node at (2.25,1.8) {\tiny GW};
            \node at (1,0.3) {\textbf{\scriptsize BZ3}};
            \node at (1.8,0.3) {\textbf{\scriptsize 5\euro} };
            \node at (0.5,4) {\textbf{\scriptsize BZ1}};
            \node at (0.5,3.6) {\textbf{\scriptsize 10\euro} };
            \node at (4,4) {\textbf{\scriptsize BZ2}};
            \node at (4,3.6) {\textbf{\scriptsize -5\euro} };
            \node[anchor=west] at (2.6,2) {\textbf{\scriptsize OBZ}};
            \node[anchor=west] at (2.6,1.7) {\textbf{\scriptsize 5\euro} };
            
            %% CAPTION
            \node[draw=none, fill=none] at (2.25,5) {\small b) Dynamic High-Price Market};
            \node[draw=none, fill=none] at (2.25,-0.5) {\small d) Offshore Bidding Zone};
            
            %% TOP-RIGHT
            \fill[fill=blueSETUP, draw=none, fill opacity=0.2] (0,5.5) -- (4.5,5.5) -- (4.5,6.1) -- (2.25,8.2) -- (0,6.7) -- cycle;
            \fill[fill=greenSETUP, draw=none, fill opacity=0.2] (4.5,6.1) -- (4.5,10) -- (2.25,10) -- (2.25,8.2) -- cycle;
            \fill[fill=yellowSETUP, draw=none, fill opacity=0.2] (0,6.7) -- (2.25,8.2) -- (2.25,10) -- (0,10) -- cycle;
            \draw[draw=white, line width = 1] (0,6.7) -- (2.25,8.2) -- (2.25,10);
            \draw[draw=white, line width = 1] (4.5,6.1) -- (2.25,8.2);
            \draw[draw=orangeSETUP, line width=1.5] (4,9) -- (2.25,7.4);
            \draw[draw=yellowSETUP!60, line width=1.5] (0.5,9.5) -- (2.25,7.4);
            \draw[draw=orangeSETUP, line width=1.5] (2,5.8) -- (2.25,7.4);
            \draw[fill=yellowSETUP, draw=black, line width = 0.5] (0,8.5) arc[start angle = 270, end angle = 360, radius=1.5] (1.5,10) -- (0,10) -- (0,8.5);
            \draw[fill=greenSETUP, draw=black, line width = 0.5] (4.5,8) -- (4.5,10) -- (3,10) arc[start angle = 192, end angle = 241.4, radius=3] -- (4.5,8);
            \fill[fill=yellowSETUP, draw=black,line width = 0.5, radius=0.4] (2.25,7.4) circle;
            \fill[fill=none, draw=black,line width = 0.5, radius=0.32] (2.25,7.4) circle;
            \draw[fill=blueSETUP, draw=black, line width = 0.5] (0,5.5) -- (3.5,5.5) arc[start angle = 0, end angle = 75, radius=0.6] arc[start angle = 75, end angle = 100.2, radius = 7] -- cycle;
            \draw[fill=none, draw=black, line width=0.5] (0,5.5) rectangle (4.5,10);
            \node[anchor=east, fill=orangeSETUP!90, draw=orangeSETUP, rotate=42, inner sep=1.2pt] at (3.16,8.55) {\textcolor{darkblueSETUP}{\scriptsize 0\tiny GW}};
            \node[anchor=west, fill=orangeSETUP!90, draw=orangeSETUP, rotate=82, inner sep=1.2pt] at (2.32,6.3) {\scalebox{0.95}{\textcolor{darkblueSETUP}{\scriptsize 0\tiny GW}}};
            \node[anchor=east, fill=yellowSETUP!70, draw=yellowSETUP, signal, signal to=west, rotate=-50, inner sep=1.2pt] at (1.48,7.94) {\textcolor{darkblueSETUP}{\scriptsize 4\tiny GW}};
            
            \node at (2.25,7.5) {\textbf{\scriptsize 4/6} };
            \node at (2.25,7.3) {\tiny GW}; 
            \node at (1,5.8) {\textbf{\scriptsize BZ3}};
            \node at (1.8,5.8) {\textbf{\scriptsize 5\euro} };
            \node at (0.5,9.5) {\textbf{\scriptsize BZ1}};
            \node at (0.5,9.1) {\textbf{\scriptsize 10\euro} };
            \node at (4,9.5) {\textbf{\scriptsize BZ2}};
            \node at (4,9.1) {\textbf{\scriptsize -5\euro} };
            
        \end{tikzpicture}
    }\vspace{-0.5em}
    \caption{Different market setup for an energy island.}
    \label{fig:setup}
    % \vspace{-1.3em}
\end{figure}
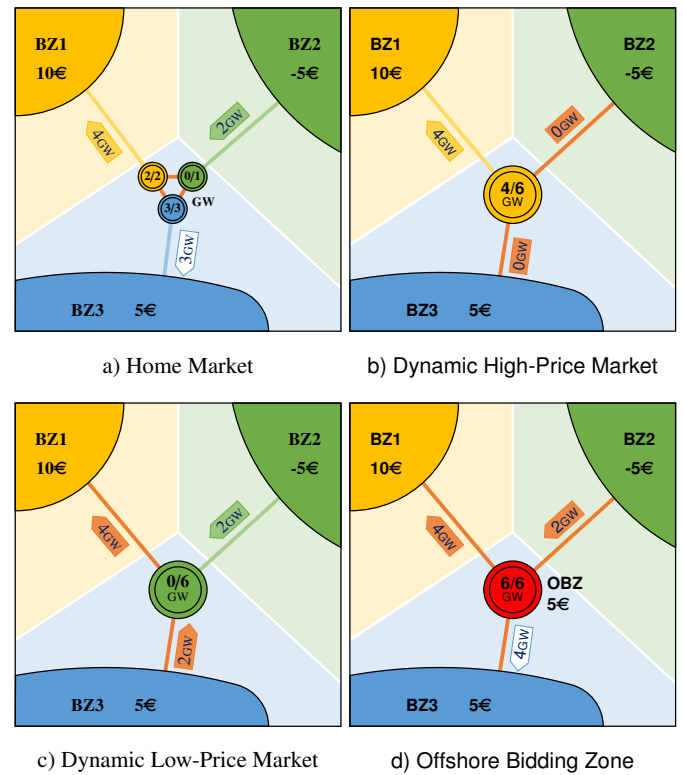
The way energy islands will be integrated to the European markets plays a deciding role in the impact the energy islands have on the social welfare of European countries and the power system operation. As it is still unclear which market setup option will be selected, this section reviews the proposed options and motivates the choice of a new offshore bidding zone in the analyses presented in this paper. The options proposed in \citet{RolandBerger} are:
\begin{enumerate}
    \item commercial flows to the respective home markets (HM);
    \item dynamic commercial flows to the high-price market (DHPM);
    \item dynamic commercial flows to the low-price market (DLPM);
    \item dedicated offshore bidding zone (OBZ).
\end{enumerate}

The following example, inspired by the \citet{NSWPHmarket} and illustrated in \figurename~\ref{fig:setup}, presents an ``extreme'' case with negative prices to highlight the differences between the four arrangements.

The system under consideration consists of 3 bidding zones (BZ) which are now connected through the hub. The three areas have respectively 4, 2 and 6 GW of installed offshore wind and transmission capacity (for a total of 12 GW). BZ1 is the high-price market (10 \euro/MWh), BZ2 the low-price market (-5 \euro/MWh) and BZ3 the medium-price market (5 \euro/MWh), and for simplicity it is assumed that these prices will not be affected by the hub. In all four situations, the wind power generated is half of the installed capacity (in total 6 GW) and wind farms have priority access to the offshore transmission capacity, meaning that exchanges between price zones are allowed only if there is some remaining transmission capacity. In \figurename~\ref{fig:setup}, the interconnectors are marked in orange, while the internal lines are marked with the color of the bidding zone. Moreover, uncongested lines have white flow indicators.

With the first option, HM, the wind farms bid into the bidding zone corresponding to their home markets and receive the corresponding electricity prices. Since priority is given to the wind farms, the imports allowed by each TSO are respectively 2GW, 1GW and 3GW. However, because BZ2 has negative price, the wind farms will not be dispatched (assuming their marginal cost is zero). The other wind farms produce respectively 2 and 3 GW (as shown inside the circles, in total 5 GW). With the remaining transmission capacity, BZ1 imports 2GW from BZ2, creating congestions on the links and confirming the price differences.

Options 2 and 3, DHPM and DLPM respectively, are conceptually similar. With the wind farms bidding to the high-price market, the hub is included in BZ1. As a consequence, the 4 GW of transmission capacity are fully utilized for the wind energy, the wind farms are dispatched only for 4 GW and no exchanges are allowed. In case of DLPM, the hub is included in BZ2, where the wind farms bid into the low-price market. Now, the wind farms are not dispatched since the price is lower than their bids. Thus, the 4 GW imported by BZ1 come from BZ2 and BZ3. Note that BZ2 can accept only 2 GW (the capacity of the link to the hub), thus the 2 GW flowing from BZ3 to the hub create a market congestion. 

Finally, the fourth option is the creation of an offshore bidding zone, which implies that all the offshore wind farms bid into a new bidding zone and are dispatched based on the market coupling with the connected zones. In this situation, all the 6 GW of wind are dispatched, and the price at the hub is equal to the price of the bidding zone with an uncongested transmission link (in the example BZ3). 

It is clear that the first three options introduce inefficiencies in the market outcome that are not present when the offshore bidding zone is introduced. Moreover, DHPM and DLPM do not comply with the Capacity Allocation and Congestion Management (CACM) Guideline, defined by the European Commission, as they are not robust and stable over time. Indeed, the low-price and high-price markets could theoretically change every hour, resulting in complicated market strategies for the market participants (mainly offshore, but also onshore). In the \citet{NSWPHmarket}, options 1 and 4 are further compared with respect to (i) compliance with European Regulations, (ii) price formation, (iii) distribution of the benefits, (iv) impact on wind farm revenues and (v) impact on balancing and operational mechanisms. Although the NSWPH members have not identified yet what is the preferable option, it seems the OBZ solution is likely to be chosen as HM requires significant regulatory changes and leads to lower socio-economic welfare. In the remainder of this paper, therefore, we consider the formation of a new offshore bidding zone as the market setup for the energy island.

% \vspace{-0.5em}
\section{Models and simulation setup}\label{sec:3}
The model of the European electricity market presented in this paper is derived from a detailed model of the pan-European transmission system. The transmission model uses as a basis a previous proposal described in \citet{BESTPATH1, BESTPATH2}, which has now been validated, extended and improved to include more detailed representations of the Nordic countries, Ireland and the UK. This section is organized in five subsections that present (i) the improved European transmission network model, (ii) the corresponding European market model, (iii) the assumptions regarding the market clearing process, (iv) the projection of the market dispatch on the transmission model and (v) the case studies. 

\subsection{European transmission network model}
\color{black}
The transmission model presented in this paper uses as a starting point the network model presented in \citet{BESTPATH2}. Originally, the model accounted for around 21'000 buses (three voltage levels: 132-150, 220 and 380-400 kV) and more than 27'000 transmission assets (AC and DC lines and 2-3 winding transformers). A set of grid reinforcement projects and new transmission lines from \citet{EHIGHWAY}, the 2020 Ten-Year Network Development Plan (TYNDP) by \citet{TYNDP2020} and the Projects of Common Interest (PCI) was included to account for the grid modifications expected by 2030. To improve the tractability of the dataset, the resulting model was reduced to comprise only the 220 and 380/400 kV levels in \citet{BESTPATH1}. The original generator and load data has been adjusted to reflect the projections for 2030 made by ENTSOE in their 2022 TYNDP \citep{TYNDP2022}. Load and RES generation at distribution level have been aggregated to the corresponding transmission substation and adjusted to 2030 projections. 

The transmission network of the Nordic countries, namely Denmark, Norway, Sweden and Finland, is taken from \citet{TOSATTO2021}. The grid reinforcements and new transmission lines presented in \citep{TYNDP2020} have been included in the grid. In addition, generator data has been modified according to \citet{TYNDP2022}, considering the phase-out of several coal-fired power plants in Denmark \citep{DANISHMIN} and Finland \citep{FINNISHMIN} and the decommissioning of nuclear power plants in Sweden \citep{SWEDISHMIN}. Load consumption and RES penetration have been adjusted based on the 2022 TYNDP of \citet{TYNDP2022}.

The data of the transmission systems of Ireland and the United Kingdom is publicly available \citep{NATIONALGRID, EIRGRID}. The Irish Transmission Statement provides detailed information regarding the modifications foreseen for 2027 in terms of transmission, generation and consumption. Generator data for the UK has been taken from \citet{ENTSOETP} and adjusted according to the UK Energy Policy \citep{UKPARLIAMENT}. Also for Ireland and the UK, demand and RES penetration have been adjusted based on the 2022 TYNDP of \citet{TYNDP2022}.

The resulting transmission network model is depicted in \figurename~\ref{fig:mapfull}. We use this model, with the inclusion of energy islands in the North Sea, to perform feasibility studies of large offshore wind installations in the North Sea region.

\definecolor{blue_map}{rgb}{0.2863,0.3294,0.4471}
\begin{figure}[!t]
    \centering
    \resizebox{0.48\textwidth}{!}{%
    \begin{tikzpicture}
        % help grid
        % \draw[help lines,step=.2] (0,0) grid (9.3,8);
        % \draw[help lines,line width=.6pt,step=1] (0,0) grid (9.3,8);
        
        \node[inner sep=0pt, anchor = south west] (network) at (0,0) {\includegraphics[trim = 2.5cm 2.5cm 18cm 2.5cm,clip,width=0.50\textwidth]{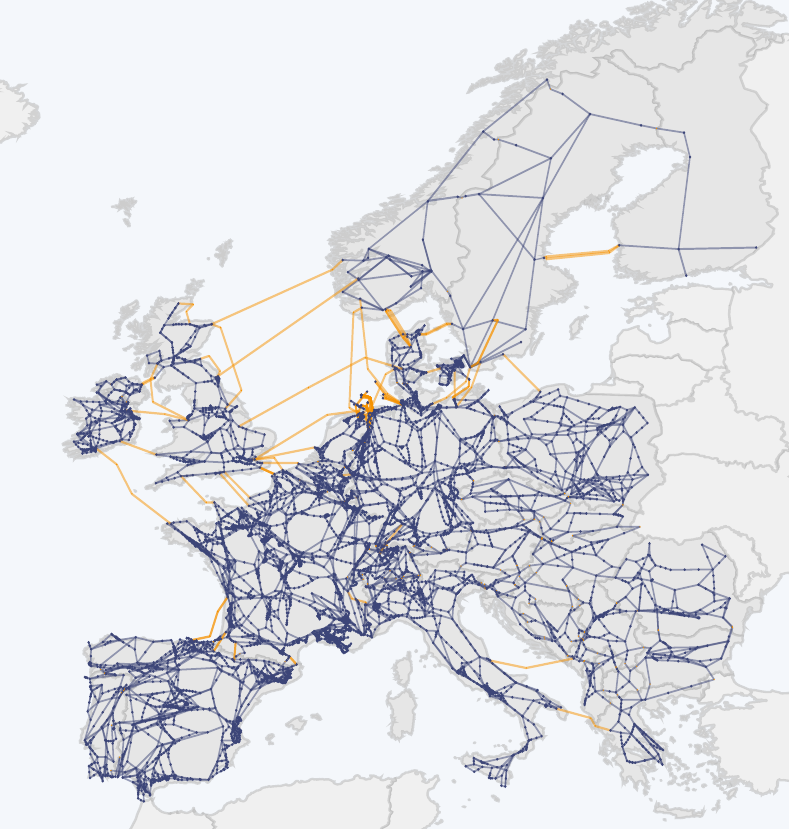}};
        \fill[fill=none, draw=blue_map, line width=1] (0,0) rectangle (9.18,9.65);

    \end{tikzpicture}
    }\vspace{-0.5em}
    \caption{European 400 kV transmission model.}
    \label{fig:mapfull}
    \vspace{-1.3em}
\end{figure}

\subsection{European electricity market model}\label{sec:marketmodel}
In Europe, a zonal pricing scheme for electricity is applied: consumers and producers within the same price zone, or bidding zone, receive the same electricity price regardless of their specific locations. Usually, a bidding zone is geographically identified with a country. However, countries can be divided into more bidding zones to reflect intra-network constraints, as in the case of Italy, Denmark, Sweden, Norway and UK. Therefore, all the nodes corresponding to one bidding zone have been aggregated to a single node, which represents the zone, obtaining an appropriate model of the European electricity market. In line with the approach adopted by ENTSO-E, Norway has been divided into 3 areas, North, Middle and South (NO-N, NO-M, NO-S), where NO-S corresponds to NO1, NO2 and NO5, NO-M to NO3, and NO-N to NO4. Ireland and Northern Ireland, instead, have been aggregated to form the Single Electricity Market \citep{IE_SEM}. As explained in Section~\ref{sec:2}, when included in the model, North Sea Energy Islands are considered as a new offshore bidding zone. 

In addition, the transmission network is included in the market model by means of equivalent interconnectors between bidding zones. The capacity of these lines are calculated based on security and reliability criteria, and TSOs often reduce the available transmission capacity to keep a certain Transmission Reliability Margin (TRM). The expected capacities for market exchanges in 2030 are taken from \citet{TYNDP2022}, which makes projections for 2030 based on new transmission projects, and are considered constant over the year.

\begin{figure}[!t]
    \centering
    \resizebox{0.48\textwidth}{!}{%
    \begin{tikzpicture}
        % help grid
        % \draw[help lines,step=.2] (0,0) grid (9.3,8);
        % \draw[help lines,line width=.6pt,step=1] (0,0) grid (9.3,8);
        
        \node[inner sep=0pt, anchor = south west] (network) at (0,0) {\includegraphics[trim = 18cm 2.5cm 2.5cm 2.5cm,clip,width=0.50\textwidth]{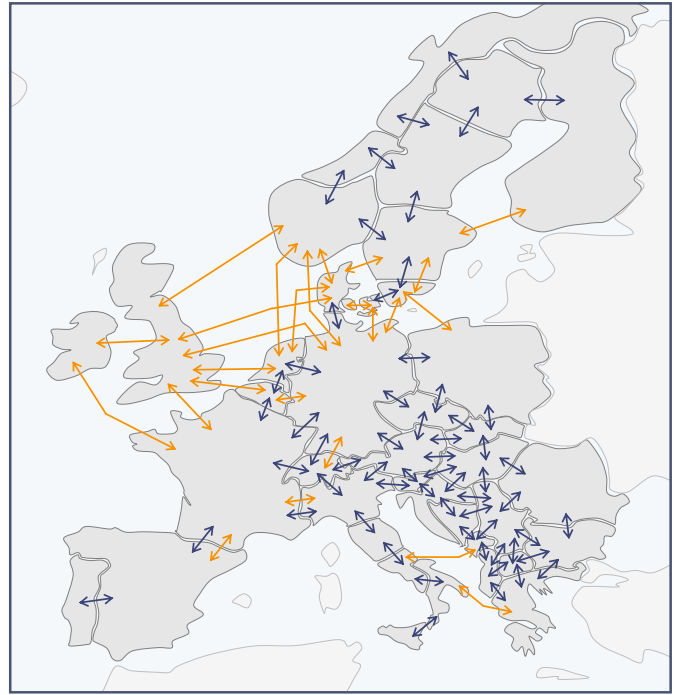}};
        \fill[fill=none, draw=blue_map, line width=1] (0,0) rectangle (9.180,9.65);

    \end{tikzpicture}
    }\vspace{-0.5em}
    \caption{European electricity market model.}
    \label{fig:mapmarket}
    % \vspace{-1.3em}
\end{figure}

Currently, European TSOs make use of two different methodologies for coupling different market zones: explicit and implicit transmission capacity auctioning \citep{NORDPOOLAUCTION}. With explicit auctions, transmission capacity and energy are traded separately in two different marketplaces. With implicit auctions, instead, different market zones are implicitly coupled and the flows on the interconnectors are the result of the energy trades. Despite being the simplest methodology to handle transmission capacity on interconnections, explicit auctioning might result in the inefficient utilization of transmission capacity, such that TSOs are gradually moving towards implicit auctioning \citep{EPEX}. Anticipating this development, we consider in our model that all countries are coupled with a single market clearing algorithm with implicit auctioning. A detailed description of the calculation of Power Transfer Distribution Factor (PTDF) matrix for flow-based market coupling is provided in \ref{sec:appendix1}.

The resulting market model is represented in \figurename~\ref{fig:mapmarket}. More specific aspects of the market clearing process will be discussed in the following. 

\subsection{Market-clearing assumptions}

The market model introduced in Section \ref{sec:marketmodel} is used for day-ahead market clearing. Ancillary services, intra-day and regulating power markets are not considered in the simulations. 

The day-ahead market clearing problem is formulated as economic dispatch with transmission grid constraints. The objective is to maximize social welfare, intended as the sum of consumer and producer surplus. The load included in the transmission model is considered inelastic, with a large value of lost load (VOLL) set to 3000 Eur/MWh. An additional 20\% of demand is considered responsive to the clearing price, with a linear utility function varying with the total inelastic demand, i.e. the slope of the line connecting 3000 Eur/MWh to 0 Eur/MWh in a range equal to 10\% of the total demand. Generator cost functions are modelled with linear coefficients which have been selected based on the fuel-type according to \citet{JENSEN2017}, and scaled-up to consider the latest projections of CO2-related costs from the 2022 TYNDP of \citet{TYNDP2022} (CO2 price is set to 70 EUR/tonne, with emission coefficients from the \citet{JRC_CO2}). We consider a market with perfect competition, meaning that all generators behave truthfully and their bids truly reflect their marginal costs of production. We do not consider subsidies from different governments and, as a result, RES participate in the market bidding at zero-marginal cost.

The constraints included in the formulation are only related to the maximum generation or consumption capacities of the market participants. No technical constraints, such as ramping limits, online and offline minimum duration periods and start-up and shut-down costs, are included in the model. Exchanges between bidding zones are calculated by means of PTDF using a linear power flow model. No transmission losses are included in the model. 

In all the simulations, the market is cleared for each hour of a time window corresponding to one year. Wind, solar and demand profiles are obtained from \citet{ENTSOSF}. For pumped-storage hydropower units in Continental Europe, limitations on water availability across the year are considered, with the results that these units participate mostly during peak-net-load hours.

For the replicability of the results, a detailed description of the market clearing algorithm is provided in \ref{sec:appendix2}.

\subsection{Projection of the market dispatch on the grid model}\label{sec:3_4}

After the market is cleared, the dispatch of generators, RES and loads is projected on the transmission model to check for its feasibility. This projection can be intended as market-based redispatching, where units are up- and down-regulated to deal with internal congestions. However, given that no reserves are procured prior to the day-ahead market clearing, this is not intended as a market analysis but as a feasibility check, and all units can be adjusted to deal with congestions. While redispatching units, the merit order curve is followed: the redispatching costs are assumed to be the absolute values of the differences between the day-ahead price and the marginal cost of production of generators. Finally, load shedding and wind curtailment are considered when necessary. For the replicability of the results, a detailed description of the projection algorithm is provided in \ref{sec:appendix2}.

\subsection{Definition of case-studies}

To study the impact of North Sea Energy Islands on the European electricity market, four different case studies have been developed with focus on different aspects, such as the size of the island, the installed transmission capacity and the connected countries. These cases cover a total of five countries facing the North Sea: Denmark, Germany, the Netherlands, Belgium and the United Kingdom. All the simulations presented in Section \ref{sec:4} are performed for a time period corresponding to one year.

\subsubsection{Reference case: no hub}
For comparison purposes, the simulation is first performed without the NESH. This case will be referred to as ``No Hub''. In the results section, most of the results with the hub will be presented as the difference with this case. 

\definecolor{blue_map}{rgb}{0.2863,0.3294,0.4471}
\definecolor{grey_map}{rgb}{0.3490,0.3490,0.3490}
\begin{figure}[!t]
    \centering
    %\resizebox{0.70\textwidth}{!}{%
    \begin{tikzpicture}
        \node[inner sep=0pt, anchor = south west] (network) at (0,0) {\includegraphics[trim = 0.05cm 9.75cm 21.34cm 0.05cm,clip,width=0.49\textwidth]{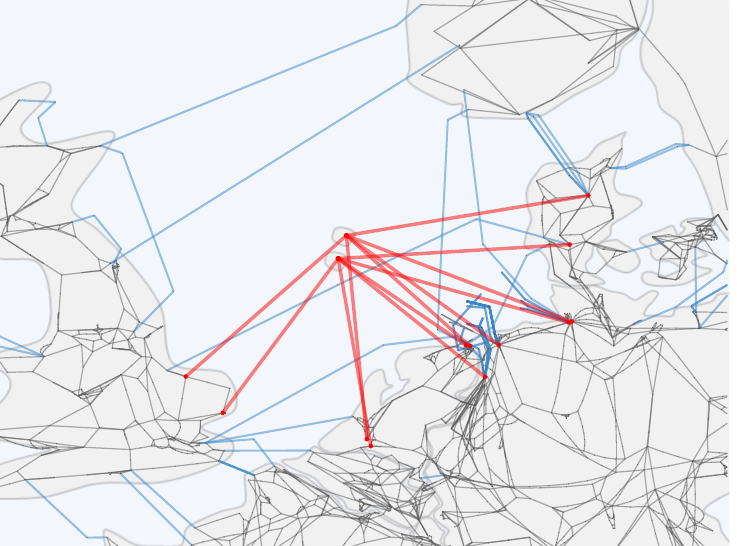}};
        \fill[fill=none, draw=blue_map, line width=1] (0,0) rectangle (8.89,6.67);
        \node[] at (3.5,3.9) {\scriptsize \textcolor{red}{\textbf{NSEH}}};
        \node[] at (1.3,2.2) {\scriptsize \textcolor{blue_map}{\textbf{UNITED}}};
        \node[] at (1.3,1.9) {\scriptsize \textcolor{blue_map}{\textbf{KINGDOM}}};
        \node[] at (6.4,5.9) {\scriptsize \textcolor{blue_map}{\textbf{NORWAY}}};
        \node[] at (7.8,3.9) {\scriptsize \textcolor{blue_map}{\textbf{DENMARK}}};
        \node[] at (7.5,1.9) {\scriptsize \textcolor{blue_map}{\textbf{GERMANY}}};
        \node[] at (3.5,0.3) {\scriptsize \textcolor{blue_map}{\textbf{FRANCE}}};
        \node[] at (4.4,0.9) {\scriptsize \textcolor{blue_map}{\textbf{BELGIUM}}};
        \node[] at (5.1,1.8) {\scriptsize \textcolor{blue_map}{\textbf{THE}}};
        \node[] at (5.6,1.5) {\scriptsize \textcolor{blue_map}{\textbf{NETHERLANDS}}};
        \node[] at (8.25,5.5) {\scriptsize \textcolor{blue_map}{\textbf{SWEDEN}}};
        \draw[very thick,red, opacity=0.5] (4.1,1.1) -- (4.13,3.53);
        \node at (4.1,1.1)[circle,fill=red,scale=0.17]{};
        % help grid
        % \draw[help lines,step=.2] (0,0) grid (8.9,6.5);
        % \draw[help lines,line width=.6pt,step=1] (0,0) grid (8.9,6.5);
    \end{tikzpicture}%}
    \caption{Connections between the NSWPH and the onshore grids. Each island has 10 GW of installed wind capacity. The number of HVDC links varies across scenario, the capacity of the each link is 1700 MW.}
\label{fig:NSWPH_map}
% \vspace{-0.5em}
\end{figure}
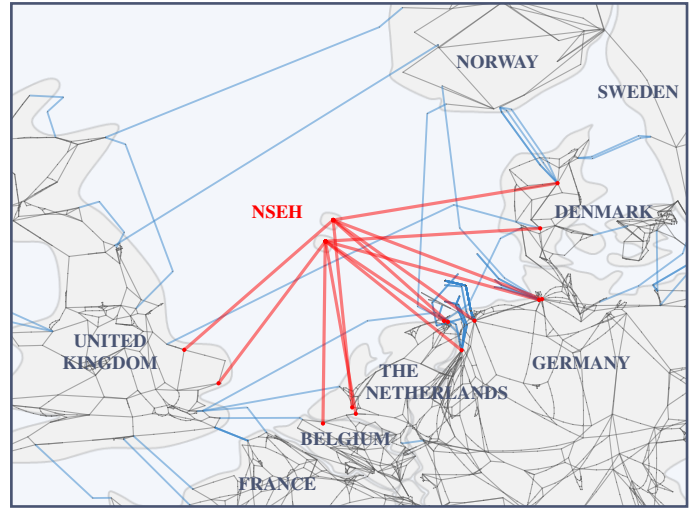

\definecolor{yellow_plot}{rgb}{1,0.752,0}
\definecolor{red_plot}{rgb}{0.874,0.325,0.152}
\definecolor{blue_plot}{rgb}{0.3020,0.7451,0.9333}
\definecolor{green_plot}{rgb}{0.572,0.815,0.313}
\definecolor{orange_plot}{rgb}{0.929,0.490,0.192}

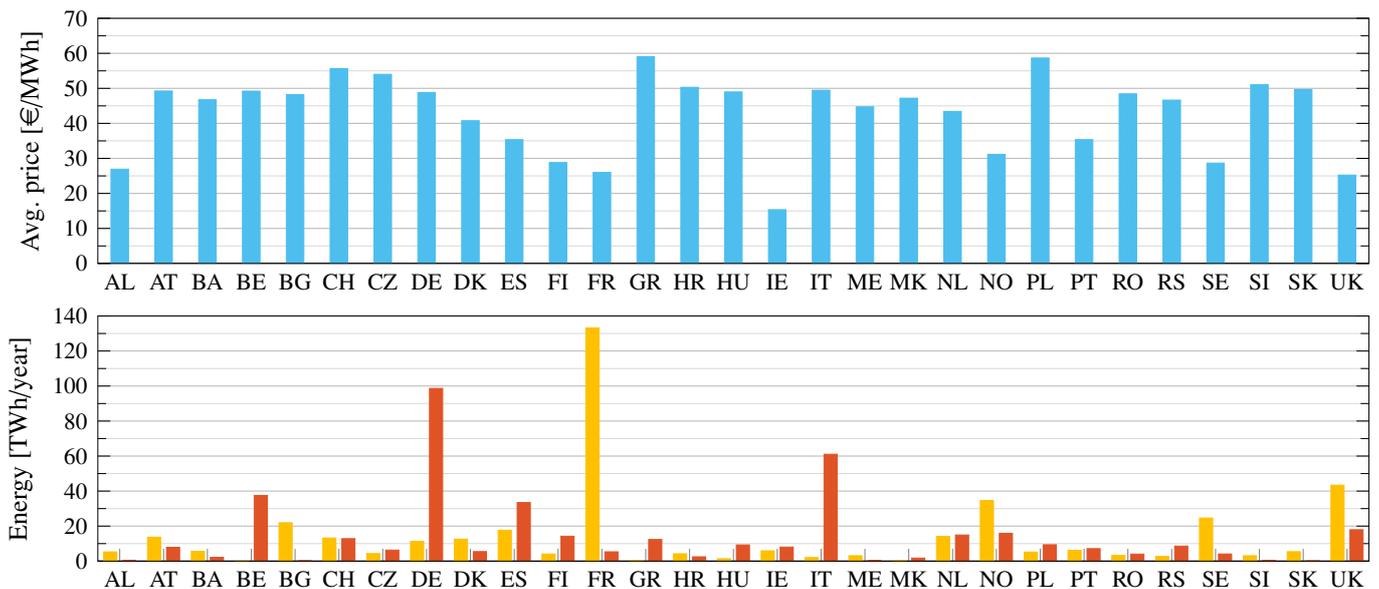
\begin{figure*}[!b]
    \centering
    \resizebox{0.991\textwidth}{!}{%
    \begin{tikzpicture}
        \begin{axis}[
            width  = 0.92\textwidth,
            height = 0.190001\textheight,
            at={(0,850)},
            bar shift auto,
            ylabel = {Avg. price [\euro/MWh]},
            ylabel style = {font=\color{black}\small},
            yticklabel style = {font=\color{black}\footnotesize},
            xticklabel style = {font=\color{black}\footnotesize},
            ytick style={black},
            y axis line  style={black,line width=1.0pt},
            xtick pos=left,
            xmin=0.5,xmax=29.5,
            xtick = {1,2,3,4,5,6,7,8,9,10,11,12,13,14,15,16,17,18,19,20,21,22,23,24,25,26,27,28,29},
            xticklabels={AL,AT,BA,BE,BG,CH,CZ,DE,DK,ES,FI,FR,GR,HR,HU,IE,IT,ME,MK,NL,NO,PL,PT,RO,RS,SE,SI,SK,UK},
            ytick = {0,10,20,30,40,50,60,70},
            ymin=0,ymax=70,
            axis line style = {line width=0.1},
            minor y tick num=1,
            ymajorgrids,
            yminorgrids,
            legend columns=4,
            legend style={at={(0.07,1)}, anchor=south west, legend cell align=left, align=left, draw=none, font=\small},
            every axis legend/.append style={column sep=0.3em},
            legend image code/.code={
                \draw [/tikz/.cd,bar width=5pt,yshift=-0.2em,bar shift=0pt]
                plot coordinates {(0cm,0.5em)};
            },
            ]
            \addplot[ybar, bar width=0.4, fill=blue_plot, draw=blue_plot, line width=0.05pt] table[row sep=crcr] {%
                1	26.93	\\
                2	49.25	\\
                3	46.75	\\
                4	49.22	\\
                5	48.21	\\
                6	55.63	\\
                7	54.01	\\
                8	48.83	\\
                9	40.76	\\
                10	35.36	\\
                11	28.82	\\
                12	25.99	\\
                13	59.03	\\
                14	50.30	\\
                15	48.99	\\
                16	15.36	\\
                17	49.48	\\
                18	44.74	\\
                19	47.20	\\
                20	43.42	\\
                21	31.16	\\
                22	58.67	\\
                23	35.36	\\
                24	48.44	\\
                25	46.65	\\
                26	28.66	\\
                27	51.04	\\
                28	49.65	\\
                29	25.24	\\
                };
            \addplot[draw=black,line width=0.3, forget plot] table[row sep=crcr] {%
                0.5 0 \\
                29.5 0 \\
            };
                
        \end{axis}%
        \begin{axis}[
            width  = 0.92\textwidth,
            height = 0.190001\textheight,
            at={(0,0)},
            bar shift auto,
            ylabel = {Energy [TWh/year]},
            ylabel style = {font=\color{black}\small},
            yticklabel style = {font=\color{black}\footnotesize},
            xticklabel style = {font=\color{black}\footnotesize},
            ytick style={black},
            y axis line  style={black,line width=1.0pt},
            xtick pos=left,
            xmin=0.5,xmax=29.5,
            xtick = {1,2,3,4,5,6,7,8,9,10,11,12,13,14,15,16,17,18,19,20,21,22,23,24,25,26,27,28,29},
            xticklabels={AL,AT,BA,BE,BG,CH,CZ,DE,DK,ES,FI,FR,GR,HR,HU,IE,IT,ME,MK,NL,NO,PL,PT,RO,RS,SE,SI,SK,UK},
            ytick = {0,20,40,60,80,100,120,140},
            ymin=0,ymax=140,
            axis line style = {line width=0.1},
            minor y tick num=1,
            ymajorgrids,
            yminorgrids,
            legend columns=4,
            legend style={at={(0.07,1)}, anchor=south west, legend cell align=left, align=left, draw=none, font=\small},
            every axis legend/.append style={column sep=0.3em},
            legend image code/.code={
                \draw [/tikz/.cd,bar width=5pt,yshift=-0.2em,bar shift=0pt]
                plot coordinates {(0cm,0.5em)};
            },
            ]
            \addplot[ybar, bar width=0.3, fill=yellow_plot, draw=yellow_plot, line width=0.05pt] table[row sep=crcr] {%
                1	5.27	\\
                2	13.66	\\
                3	5.60	\\
                4	0.07	\\
                5	21.92	\\
                6	13.21	\\
                7	4.32	\\
                8	11.31	\\
                9	12.58	\\
                10	17.65	\\
                11	4.06	\\
                12	133.14	\\
                13	0.25	\\
                14	4.20	\\
                15	1.37	\\
                16	5.90	\\
                17	2.04	\\
                18	3.09	\\
                19	0.37	\\
                20	14.14	\\
                21	34.63	\\
                22	5.18	\\
                23	6.15	\\
                24	3.26	\\
                25	2.68	\\
                26	24.61	\\
                27	3.08	\\
                28	5.39	\\
                29	43.37	\\  
            };
            \addplot[ybar, bar width=0.3, fill=red_plot, draw=red_plot, line width=0.05pt] table[row sep=crcr] {%
                1	0.51	\\
                2	7.89	\\
                3	2.17	\\
                4	37.53	\\
                5	0.25	\\
                6	12.83	\\
                7	6.30	\\
                8	98.65	\\
                9	5.44	\\
                10	33.45	\\
                11	14.23	\\
                12	5.32	\\
                13	12.36	\\
                14	2.47	\\
                15	9.26	\\
                16	8.03	\\
                17	61.04	\\
                18	0.43	\\
                19	1.75	\\
                20	14.92	\\
                21	15.90	\\
                22	9.35	\\
                23	7.13	\\
                24	4.02	\\
                25	8.58	\\
                26	4.09	\\
                27	0.49	\\
                28	0.18	\\
                29	17.93	\\  
            };
            
            \addplot[draw=black,line width=0.3, forget plot] table[row sep=crcr] {%
                0.5 0 \\
                29.5 0 \\
            };
                
        \end{axis}%
    \end{tikzpicture}
    }%
    \vspace{-0.5em}
    \caption{Average prices (upper figure) and imports/exports (lower figure) of each European country in the reference case.}
    \label{fig:primpexp_REF}
    % \vspace{-1em}
\end{figure*}
\subsubsection{10 GW and 20 GW islands}
The size of the hub is defined by the amount of installed wind capacity. Two different sizes are simulated: 10~GW and 20~GW. In the 10~GW case, one island is built in the North Sea, and 6 HVDC links connect the island to Denmark (DK1), Germany, the Netherlands and Belgium. Each link has 1700~MW transfer capacity; Germany and the Netherlands have two links each. In the 20~GW case, a second island is considered. The second island is an exact copy of the first, with 10~GW of wind capacity installed and 6 HVDC links with 1700~MW transfer capability; however, the connection points to the onshore grids have been changed. \figurename~\ref{fig:NSWPH_map} shows the configuration with two islands.

\subsubsection{More transmission capacity for exchanges}
Since the construction of the hub requires a certain transmission capacity, this case study (``Exchanges'') investigates the impact of increasing this capacity for allowing more exchanges through the hub. This case is similar to the ``10 GW'' case, except for the total transmission capacity which is increased to 15 GW (instead of 10 GW). The transfer capacity from the hub to Denmark and Belgium is 2.5~GW (for each country), and to Germany and the Netherlands 5~GW (for each country).

\subsubsection{Connection to the UK}
This case study (``UK'') aims at investigating how a new interconnection, in this case to the UK, would alter the equilibrium of the NSEH-connected countries. As in the previous case, only one island is built with 10~GW of wind power. The 10~GW of transmission capacity are now distributed to five countries. Germany and the Netherlands have a total of 2.9~GW of transmission capacity each, while Denmark, Belgium and the UK have 1.45~GW each. 

% \vspace{-0.5em}
\section{Results and discussion}\label{sec:4}
In this section, the market model presented in Section \ref{sec:marketmodel} has been used to evaluate the impact of different combinations of installed wind and transmission capacity in the North Sea on the European countries. The structure of this section is the following. First, the results obtained for the reference case are presented so that the reader can get familiar with the model. The discussion will then focus on certain attributes of interest, such as electricity prices, exchanges between countries, and utilization of transmission assets for the countries directly connected to the energy island. Subsequently, the analyses will move towards the other European countries to highlight the repercussions of hybrid projects in the North Sea on zones not directly involved in the projects. The potential revenues of wind power producers and the transmission system operator on the island are then considered. Finally, the discussion will focus on system needs and market challenges that could arise with these projects, looking at changes of behavior of market players and at the technical feasibility of such projects. For transparency, \ref{sec:appendix3} presents the numerical values of average electricity prices, total generation, consumption, exports and imports for all the bidding zones across the five scenarios.

\subsection{Reference case}

The reference case considers the European electricity market in 2030 without any energy island in the North Sea. 

The first attribute to be considered is the distribution of electricity prices across the continent. \figurename~\ref{fig:primpexp_REF} (upper figure) shows the basedload electricity price per country. For those countries with multiple bidding zones, the price is weighted on the volumes traded in each price area. Note that DE corresponds to Germany and Luxembourg unless otherwise stated, and IE to Ireland and Northern Ireland. The countries with the lowest price are Ireland and the UK, where the aggressive RES build-out pushes power prices down, France, and hydropower-dominated markets such as Sweden or Albania. On the other side, Greece is the country with the highest price, followed by Poland, Switzerland, Czech Republic and Italy.

The distribution of electricity prices is also reflected on the net position of each country. \figurename~\ref{fig:primpexp_REF} (lower figure) shows the imports/exports of each country. The energy that France exports stands out compared to the other countries; France has a large amount of installed nuclear and RES capacity, with exports equal to 20\% of the total production. Second in the list is the UK, followed by Sweden, Bulgaria and Norway. 

Concerning the imports, Germany imports almost 20\% of its total consumption: because of the nuclear and coal/lignite phase-out, large shares of energy are imported during low-wind periods. The second largest importer is Italy, with 25\% of its demand coming from France, Switzerland and Eastern Europe. Third in terms of volume is Belgium, but its share of imported energy corresponds to 65\% of the total consumption, the highest in terms of percentage. Similar to Germany, the nuclear phase-out in Belgium leads to high imports during low wind periods. 
\color{black}

\input{Plots/plot_congREF}

Finally, \figurename~\ref{fig:congREF} integrates the information related to prices and exchanges with the main directions of power flows and the occurrence of congestions in all corridors. The first observation is that power flows from Northern, Eastern and Western Europe to Central and Southern Europe. Second, the full controllability of HVDC lines helps bypass congestions in AC corridors, as HVDC flows are not dependent on the impedance ratios of the parallel AC corridors. This is particularly clear looking at the case of the UK: the total incoming flows amount to 75 TWh, but the total imports are only 44 TWh. The remaining 31 TWh are redirected towards Belgium, Netherlands, Germany, Denmark, Ireland and Northern Ireland. These loop-flows over the HVDC connections from France to the UK and back to Continental Europe occur as they help increase the exchanges between France and Belgium, since the AC corridor is congested for more than 60\% of the time. Similar situations are encountered in Denmark, the Netherlands, Germany and, more in general, in all the zones that are close to high price markets. The last observation is related to the correlation between prices and congestions. It can be noticed from \figurename~\ref{fig:congREF} that the congestions always occur between the zones with the lowest and highest prices. Or better, low and high prices are caused by congestions, with the zones at the two ends of the congested corridor experiencing the lowest and highest prices \citep{OREN1995}. This happens for example between France and Belgium, Albania and Greece,  and so on. 

\subsection{Impact on the countries connected to the hub}
\definecolor{orange_plot}{rgb}{0.964,0.572,0}
\definecolor{red_plot}{rgb}{0.874,0.325,0.152}
\definecolor{blue_plot}{rgb}{0,0.439,0.752}
\definecolor{green_plot}{rgb}{0.572,0.815,0.313}

\definecolor{blue_plot1}{rgb}{0.109,0.6784,0.8941}
\definecolor{blue_plot2}{rgb}{0.1490,0.5137,0.7765}
\definecolor{green_plot1}{rgb}{0.2588,0.7294,0.5922}
\definecolor{green_plot2}{rgb}{0.2431,0.5333,0.3255}

\definecolor{red_plot1}{rgb}{0.9098,0.2980,0.1333}%
\definecolor{orange_plot1}{rgb}{1.0000,0.7412,0.2784}%
\definecolor{red_plot2}{rgb}{0.7137,0.2863,0.1490}%
\definecolor{orange_plot2}{rgb}{1.0000,0.5176,0.1529}%

\definecolor{yellow_plot1}{rgb}{1.0000,0.9176,0.6118}%
\definecolor{brown_plot1}{rgb}{0.8078,0.5529,0.2431}%
\definecolor{yellow_plot2}{rgb}{1.0000,0.7922,0.0314}%
\definecolor{brown_plot2}{rgb}{0.5176,0.4039,0}%

\definecolor{green2_plot}{rgb}{0.329411764705882,0.619607843137255,0.223529411764706}%
\definecolor{green3_plot}{rgb}{0.541176470588235,0.721568627450980,0.200000000000000}%
\definecolor{green1_plot}{rgb}{0.752941176470588,0.811764705882353,0.227450980392157}%
\definecolor{green4_plot}{rgb}{0.00784313725490196,0.588235294117647,0.462745098039216}%

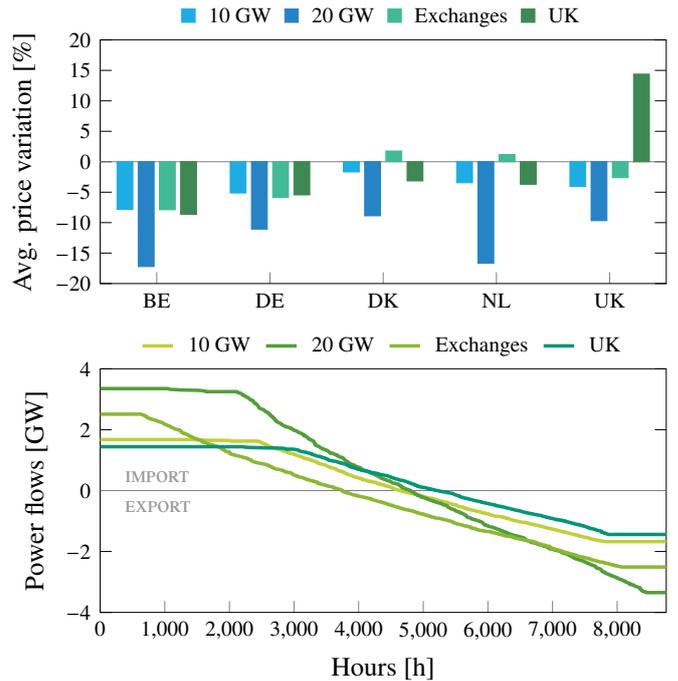
\begin{figure}[!t]
    \centering
    \resizebox{0.48\textwidth}{!}{%
    \begin{tikzpicture}
        \begin{axis}[
            anchor = south,
            width  = 0.49\textwidth,
            height = 0.200001\textheight,
            at={(0,0)},
            % axis y line*=left,
            % axis x line=box,
            % ybar=0.5pt,
            % bar width=0.25,
            % scale only axis,
            bar shift auto,
            ylabel = {Avg. price variation [\%]},
            ylabel style = {font=\color{black}},
            yticklabel style = {font=\color{black}\footnotesize},
            xticklabel style = {font=\color{black}\footnotesize},
            ytick style={black},
            y axis line  style={black,line width=1.0pt},
            xtick pos=left,
            xmin=0.5,xmax=5.5,
            xtick = {1,2,3,4,5},
            xticklabels={{BE},{DE},{DK},{NL},{UK}},
            ytick = {-20,-15,-10,-5,0,5,10,15,20},
            ymin=-20,ymax=20,
            axis line style = {line width=0.1},
            legend columns=4,
            legend style={at={(0.5,1.01)}, anchor=south, legend cell align=left, align=left, draw=none, font=\footnotesize},
            every axis legend/.append style={column sep=0.3em},
            legend image code/.code={
                \draw [/tikz/.cd,bar width=5pt,yshift=-0.2em,bar shift=0pt]
                plot coordinates {(0cm,0.5em)};
            },
            ]
            \addplot[draw=black!50, forget plot, line width = 0.3] table[row sep=crcr] {%
                0   0\\
                6   0\\
                };
            \addplot[ybar, bar width=0.14, fill=blue_plot1, draw=blue_plot1, line width=0.05pt] table[row sep=crcr] {%
                1	-7.838068068	\\
                2	-5.129971736	\\
                3	-1.673593519	\\
                4	-3.452852342	\\
                5	-4.043347115	\\
                };
            \addplot[ybar, bar width=0.14, fill=blue_plot2, draw=blue_plot2, line width=0.05pt] table[row sep=crcr] {%
                1	-17.21469446	\\
                2	-11.06093606	\\
                3	-8.878730821	\\
                4	-16.682742	\\
                5	-9.673892649	\\
                };
            \addplot[ybar, bar width=0.14, fill=green_plot1, draw=green_plot1, line width=0.05pt] table[row sep=crcr] {%
                1	-7.881934313	\\
                2	-5.84870593	\\
                3	1.761508576	\\
                4	1.181625613	\\
                5	-2.617851247	\\
                };
            \addplot[ybar, bar width=0.14, fill=green_plot2, draw=green_plot2, line width=0.05pt] table[row sep=crcr] {%
                1	-8.63966955	\\
                2	-5.440201481	\\
                3	-3.149907015	\\
                4	-3.697476388	\\
                5	14.39018505	\\
                };
            \legend{10 GW, 20 GW, Exchanges, UK}
        \end{axis}%
        \begin{axis}[
            anchor = north,
            width  = 0.49\textwidth,
            height = 0.200001\textheight,
            at={(0,-280)},
            % axis y line*=left,
            % axis x line=box,
            % ybar=0.5pt,
            % bar width=0.25,
            % scale only axis,
            ylabel = {Power flows [GW]},
            xlabel = {Hours [h]},
            ylabel style = {font=\color{black}},
            yticklabel style = {font=\color{black}\footnotesize},
            xticklabel style = {font=\color{black}\footnotesize},
            ytick style={black},
            y axis line  style={black,line width=1.0pt},
            xtick pos=left,
            xmin=0,xmax=8760,
            ymin=-4,ymax=4,
            axis line style = {line width=0.1},
            legend columns=4,
            legend style={at={(0.5,1.01)}, anchor=south, legend cell align=left, align=left, draw=none, font=\footnotesize},
            every axis legend/.append style={column sep=0.3em},
            legend image code/.code={
                \draw[mark repeat=2,mark phase=2]
                plot coordinates {
                    (0cm,0cm)
                    (0.2cm,0cm)        %% default is (0.3cm,0cm)
                    (0.4cm,0cm)         %% default is (0.6cm,0cm)
                    };
                }
            ]
            \addplot[draw=black!50, forget plot, line width = 0.3] table[row sep=crcr] {%
                0   0\\
                8760   0\\
                };
            \addplot[draw=green1_plot, line width=1.3pt] table[y=$s1$, x=$h$]{Plots/Data/flowsNSWPH_BE.dat};
            \addplot[draw=green2_plot, line width=1.3pt] table[y=$s2$, x=$h$]{Plots/Data/flowsNSWPH_BE.dat};
            \addplot[draw=green3_plot, line width=1.3pt] table[y=$s3$, x=$h$]{Plots/Data/flowsNSWPH_BE.dat};
            \addplot[draw=green4_plot, line width=1.3pt] table[y=$s4$, x=$h$]{Plots/Data/flowsNSWPH_BE.dat};
            \legend{10 GW, 20 GW, Exchanges, UK}
        \end{axis}
        \node[anchor=south west] at (-3.5,-2.75) {\small \textcolor{gray!80}{\textsc{import}}};
        \node[anchor=north west] at (-3.5,-2.74) {\small \textcolor{gray!80}{\textsc{export}}};
        
    \end{tikzpicture}
    }%
    % \vspace{-1.9em}
    \caption{Average price variation (left) and duration curve of flows on the link between DK and the hub (right) in the zones connected to the hub across the different scenarios.}
    \label{fig:price_NSEH}
    % \vspace{-1em}
\end{figure}
From this section on, the impact of different energy hub configurations are investigated; in particular, this section focuses on the countries directly connected to the island, i.e. Germany, the Netherlands, Denmark, Belgium and, in the last scenario, the UK. 

\figurename~\ref{fig:price_NSEH} (left figure) shows the percentage variation of the average prices in the five countries. Given that a large share of \mbox{zero-marginal} cost generation has been placed in the North Sea, one would expect that all the countries directly connected to the hub would face a price reduction. However, this does not always happen because, together with generation capacity, a large amount of transmission capacity is installed, facilitating further exchanges between the connected countries.

Prices in the Netherlands and in Denmark are lower than prices in Germany and Belgium. It follows that prices tend to converge to a common value as transmission capacity is installed between these countries: in the first two scenarios (10 GW, 20 GW and Exchanges), prices decrease in all countries as transmission capacity is mainly used to import wind energy. In the scenario ``Exchanges'', however, the difference between the installed wind and transmission capacity is 5 GW; the additional trades between the countries result in a further decrease of prices in DE and BE, while prices slightly increase in DK and NL. In the last scenario, instead, the low-price market connected to the hub is the UK and, thus, electricity prices decrease in all other countries, while they increase in the UK.

The trend of electricity prices finds confirmation in \figurename~\ref{fig:price_NSEH} (right figure), which shows the duration curve of the flows on the link between DK and the hub. The energy imported/exported through the hub is equal to the area defined from each curve and the x-axis. Moving from the ``10 GW'' to the ``20 GW'' scenarios, the ratio between imported and exported energy by the UK stays the same, explaining why the price variation is small. On the contrary, in the third scenario the export area increases significantly, while in the fourth it decreases, explaining the large price increase in the third scenario and the opposite trend in the fourth.

\definecolor{green2_plot}{rgb}{0.329411764705882,0.619607843137255,0.223529411764706}%
\definecolor{green3_plot}{rgb}{0.541176470588235,0.721568627450980,0.200000000000000}%
\definecolor{green1_plot}{rgb}{0.752941176470588,0.811764705882353,0.227450980392157}%
\definecolor{green4_plot}{rgb}{0.00784313725490196,0.588235294117647,0.462745098039216}%

\definecolor{yellow_plot1}{rgb}{1.0000,0.9176,0.6118}%
\definecolor{brown_plot1}{rgb}{0.8078,0.5529,0.2431}%
\definecolor{yellow_plot2}{rgb}{1.0000,0.7922,0.0314}%
\definecolor{brown_plot2}{rgb}{0.5176,0.4039,0}%

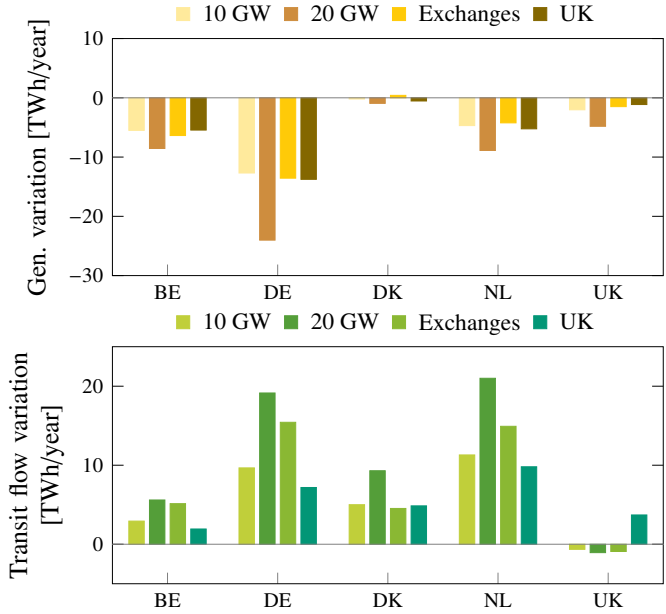
\begin{figure}[!t]
    \centering
    \resizebox{0.48\textwidth}{!}{%
    \begin{tikzpicture}
    
    \begin{axis}[
            anchor = south west,
            width  = 0.49\textwidth,
            height = 0.200001\textheight,
            at={(0,0)},
            % axis y line*=left,
            % axis x line=box,
            % ybar=0.5pt,
            % bar width=0.25,
            % scale only axis,
            bar shift auto,
            ylabel = {Gen. variation [TWh/year]},
            ylabel style = {font=\color{black}},
            yticklabel style = {font=\color{black}\footnotesize},
            xticklabel style = {font=\color{black}\footnotesize},
            ytick style={black},
            y axis line  style={black,line width=1.0pt},
            xtick pos=left,
            xmin=0.5,xmax=5.5,
            xtick = {1,2,3,4,5},
            xticklabels={{BE},{DE},{DK},{NL},{UK}},
            ytick = {-30,-20,-10,0,10},
            ymin=-30,ymax=10,
            axis line style = {line width=0.1},
            legend columns=4,
            legend style={at={(0.5,1.01)}, anchor=south, legend cell align=left, align=left, draw=none, font=\small},
            every axis legend/.append style={column sep=0.3em},
            legend image code/.code={
                \draw [/tikz/.cd,bar width=5pt,yshift=-0.2em,bar shift=0pt]
                plot coordinates {(0cm,0.5em)};
            },
            ]
            \addplot[ybar, bar width=0.14, fill=yellow_plot1, draw=yellow_plot1, line width=0.05pt] table[row sep=crcr] {%
                1	-5.498000839	\\
                2	-12.69143995	\\
                3	-0.202904574	\\
                4	-4.691478043	\\
                5	-2.025592151	\\
                };
            \addplot[ybar, bar width=0.14, fill=brown_plot1, draw=brown_plot1, line width=0.05pt] table[row sep=crcr] {%
                1	-8.548781615	\\
                2	-24.01649131	\\
                3	-0.937869001	\\
                4	-8.884558704	\\
                5	-4.805947537	\\
                };
            \addplot[ybar, bar width=0.14, fill=yellow_plot2, draw=yellow_plot2, line width=0.05pt] table[row sep=crcr] {%
                1	-6.368714171	\\
                2	-13.56100624	\\
                3	0.414161488	\\
                4	-4.22983451	\\
                5	-1.491815958	\\
                };
            \addplot[ybar, bar width=0.14, fill=brown_plot2, draw=brown_plot2, line width=0.05pt] table[row sep=crcr] {%
                1	-5.444913654	\\
                2	-13.7561328	\\
                3	-0.53679595	\\
                4	-5.238075874	\\
                5	-1.136809231	\\
                };
            \addplot[draw=black!50, forget plot, line width = 0.3] table[row sep=crcr] {%
                0   0\\
                6   0\\
                };
            \legend{10 GW, 20 GW, Exchanges, UK}
        \end{axis}

        \begin{axis}[
            anchor = north west,
            width  = 0.49\textwidth,
            height = 0.200001\textheight,
            at={(0,-90)},
            % axis y line*=left,
            % axis x line=box,
            % ybar=0.5pt,
            % bar width=0.25,
            % scale only axis,
            bar shift auto,
            ylabel = {Transit flow~\mbox{variation} [TWh/year]},
            ylabel style = {font=\color{black},align=center,text width = 4cm},
            yticklabel style = {font=\color{black}\footnotesize},
            xticklabel style = {font=\color{black}\footnotesize},
            ytick style={black},
            y axis line  style={black,line width=1.0pt},
            xtick pos=left,
            xmin=0.5,xmax=5.5,
            xtick = {1,2,3,4,5},
            xticklabels={{BE},{DE},{DK},{NL},{UK}},
            ymin=-5,ymax=25,
            axis line style = {line width=0.1},
            legend columns=4,
            legend style={at={(0.5,1.01)}, anchor=south, legend cell align=left, align=left, draw=none, font=\small},
            every axis legend/.append style={column sep=0.3em},
            legend image code/.code={
                \draw [/tikz/.cd,bar width=5pt,yshift=-0.2em,bar shift=0pt]
                plot coordinates {(0cm,0.5em)};
            },
            ]
            \addplot[ybar, bar width=0.14, fill=green1_plot, draw=green1_plot, line width=0.05pt] table[row sep=crcr] {%
                1	2.930	\\
                2	9.662	\\
                3	5.003	\\
                4	11.302	\\
                5	-0.647	\\
                };
            \addplot[ybar, bar width=0.14, fill=green2_plot, draw=green2_plot, line width=0.05pt] table[row sep=crcr] {%
                1	5.601	\\
                2	19.144	\\
                3	9.302	\\
                4	21.005	\\
                5	-1.065	\\
                };
            \addplot[ybar, bar width=0.14, fill=green3_plot, draw=green3_plot, line width=0.05pt] table[row sep=crcr] {%
                1	5.136	\\
                2	15.434	\\
                3	4.53	\\
                4	14.927	\\
                5	-0.92	\\
                };
            \addplot[ybar, bar width=0.14, fill=green4_plot, draw=green4_plot, line width=0.05pt] table[row sep=crcr] {%
                1	1.93	\\
                2	7.182	\\
                3	4.856	\\
                4	9.804	\\
                5	3.706	\\
                };
            \addplot[draw=black!50, forget plot, line width = 0.3] table[row sep=crcr] {%
                0   0\\
                6   0\\
                };
            \legend{10 GW, 20 GW, Exchanges, UK}
            \end{axis}
    \end{tikzpicture}
    }%
    % \vspace{-1.9em}
    \caption{Generation changes (left) and transit flow variation (right) in the zones connected to the hub across the different scenarios.}
    \label{fig:flow_NSEH}
    % \vspace{-1em}
\end{figure}

In terms of generation, the newly installed wind capacity is placed at the beginning of the merit order curve and shifts conventional generators to the right. This is reflected in \figurename~\ref{fig:flow_NSEH} (left graph), which shows the changes in the energy generated in the five considered countries. The largest variation is found in Germany, where the energy imported from the hub replaces domestic production (of which 80\% comes from fossil fuels). Similarly, fossil fuel generation decreases in all other countries. In DK and NL, the decrease in the third scenario is lower than in the first one because of the additional trades.

An interesting aspect that can be observed in \figurename~\ref{fig:flow_NSEH} (right graph) is that not all energy flowing from the hub to the different countries targets domestic consumption. Indeed, \figurename~\ref{fig:flow_NSEH} shows the variation of transit flows in the five considered countries. As expected, as long as there is enough transmission capacity in the AC system, power flows to those countries with the most expensive generation. For example, power flows from the Netherlands to Belgium and Germany, and from Germany to Switzerland and Austria, and so on. Transit flows often occur in AC systems and in countries with multiple HVDC connections. In the latter case, an adequate compensation mechanism for the costs of HVDC losses and losses induced in the intra-zonal network should be considered \citep{TOSATTO2020}, as the Inter-TSO Compensation (ITC) mechanism does not cover flows on HVDC links, which are considered as ``individual transactions for declared transit flows'' \citep{EU838}.

Finally, \tablename~\ref{tab1:rev} presents the variation of producer and consumer surplus. In general, with lower electricity prices, revenues of generating companies decrease, while the surplus of consumers increases. Indeed, generator surplus is calculated as the product of the energy produced times the difference between the electricity price and the cost of production. Consumer surplus, instead, is the product of the energy consumed times the difference between their utility and the electricity price. In addition, part of domestic generation is now replaced by imports from the hub, further decreasing producer surplus. As a general trend, it can be seen that social welfare, defined as the sum of consumer and producer surplus, increases in most of the scenarios. Note that, conventionally within the power system operation and electricity market literature (and practice), social welfare only relies on short-term costs and utilities, leading to consumer and producer surpluses. This means that such a definition of social welfare does not account for the costs of investments (unless the investment in production and consumption assets is reflected in the offers to the electricity markets), especially not investments at system level. In the three scenarios where the UK is not connected to the hub, the lower prices and the large amount of wind generation in the NSEH result in a loss for English generating companies which, in turn, results in a reduction of social welfare. When connected to the hub, instead, the additional transmission capacity makes it possible to increase the exports to the countries connected to the hub, resulting in a positive increase. This is experienced also in Denmark in the ``20 GW'' scenario, as the increase in wind generation reduces the revenues from international trading for Danish generators.

\begin{table}[!t]
\vspace{-0.5em}
    \caption{Producer and consumer surplus variation in the considered countries across the different scenarios (million Euros per year).}
    \label{tab1:rev}
    \small
    \centering
        \begin{tabularx}{0.48\textwidth}{*1{>{\arraybackslash}X} *5{>{\centering\arraybackslash}X}}
        \hline
        & & {\bf 10 GW} & {\bf 20 GW} & {\bf Exch.} & {\bf UK}\Tstrut\Bstrut\\ 
        \hline
        \multirow{4}{*}{\vtop{\hbox{\strut \textbf{Producer}}\hbox{\strut \textbf{Surplus}}}} &	BE	&	-176	&	-403	&	-161	&	-198            \Tstrut\\
        &	DE	&	-1315	&	-2818	&	-1513	&	\text{-1374 }   \\
        &	DK	&	-36	    &	-219	&	66	    &	\text{-66 }     \\
        &	NL	&	-2	    &	-838	&	430	    &	\text{10 }      \\
        &	UK	&	-337	&	-834	&	-215	&	1431	        \Bstrut\\

        \multirow{4}{*}{\vtop{\hbox{\strut \textbf{Consumer}}\hbox{\strut \textbf{Surplus}}}} &	BE	&	368	    &	818	    &	369	    &	402	            \Tstrut\\
        &	DE	&	1509	&	3196	&	1728	&	\text{1590 }    \\
        &	DK	&	50	    &	215	    &	-26	    &	\text{79 }      \\
        &	NL	&	179	    &	1039	&	-133	&	\text{187 }     \\
        &	UK	&	269	    &	698	    &	174	    &	-879	        \Bstrut\\
        \hline
   \end{tabularx}
\end{table}

\input{Plots/plot_revOTHERS}

\definecolor{orange_sw}{rgb}{1.00000,0.60000,0.00000}%
\definecolor{blue_sw}{rgb}{0.0314,0.2510,0.5059}%

\begin{figure*}[!b]
    \centering
    \resizebox{0.98\textwidth}{!}{%
    \begin{tikzpicture}
        \begin{axis}[
            width  = 1\textwidth,
            height = 0.220001\textheight,
            at={(0,0)},
            axis on top,
            % axis y line*=left,
            % axis x line=box,
            % ybar=0.5pt,
            % bar width=0.25,
            % scale only axis,
            xlabel = {Import/Export [TWh/year]},
            ylabel = {Social welfare variation [M\euro/year]},
            ylabel style = {font=\color{black}},
            yticklabel style = {font=\color{black}\footnotesize},
            xticklabel style = {font=\color{black}\footnotesize},
            ytick style={black},
            y axis line  style={black,line width=1.0pt},
            xtick pos=left,
            xmin=-120,xmax=140,
            ytick = {-200,-100,0,100,200,300,400,500},
            ymin=-200,ymax=500,
            axis line style = {line width=0.1},
            legend columns=3,
            legend style={at={(0.5,1.02)}, anchor=south, legend cell align=left, align=left, draw=none, font=\small},
            every axis legend/.append style={column sep=0.4em},
            ]
            \addplot [black, forget plot] table[row sep=crcr]{%
            0   -200\\
            0   500\\
            };
            \addplot [black, forget plot] table[row sep=crcr]{%
            -120   0\\
            140   0\\
            };
            \addplot [color=white, line width=1pt, draw=none, mark size=2pt, mark=o, mark options={solid, blue_sw}] table[row sep=crcr]{%
                -100.41	194.19	\\
                -61.83	50.27	\\
                -42.94	191.65	\\
                -16.08	5.94	\\
                -12.13	0.43	\\
                -10.46	1.62	\\
                -8.04	2.72	\\
                -6.23	1.67	\\
                -5.86	1.78	\\
                -5.34	176.28	\\
                -3.20	4.37	\\
                -2.24	5.15	\\
                -1.41	0.14	\\
                -1.03	-0.69	\\
                -0.97	0.60	\\
                0.00	5.11	\\
                1.60	-1.13	\\
                2.38	-1.09	\\
                2.62	-0.21	\\
                2.98	0.52	\\
                4.61	7.75	\\
                4.75	1.23	\\
                5.12	-2.29	\\
                6.90	14.12	\\
                18.74	42.28	\\
                20.02	13.89	\\
                21.55	-1.19	\\
                23.39	-68.06	\\
                124.62	-22.46	\\
            };

            \addplot [color=white, line width=1pt, draw=none, mark size=2pt, mark=x, mark options={solid, orange_sw}] table[row sep=crcr]{%
                -112.16	378.77	\\
                -64.42	82.84	\\
                -46.18	414.46	\\
                -16.39	13.82	\\
                -12.14	0.32	\\
                -10.94	4.32	\\
                -9.87	200.36	\\
                -8.18	6.06	\\
                -7.56	3.45	\\
                -6.52	3.09	\\
                -4.46	7.99	\\
                -2.36	6.75	\\
                -1.41	0.45	\\
                -1.20	1.25	\\
                -1.10	-0.35	\\
                -0.31	-0.43	\\
                1.48	-1.96	\\
                2.11	-1.89	\\
                2.56	1.08	\\
                2.57	-0.53	\\
                3.45	14.72	\\
                4.74	2.16	\\
                5.05	-4.34	\\
                6.10	-4.04	\\
                18.75	115.63	\\
                19.09	21.89	\\
                20.46	-136.02	\\
                21.39	-2.24	\\
                119.73	-72.72	\\
            };
                
            % \addplot [color=blue_sw, line width=1pt, dashed, domain=-80:220,forget plot] (x,-0.2394*x+3.1923);
            % \addplot [color=orange_sw, line width=1pt, dashed, domain=-80:220,forget plot] (x,-0.5114*x+6.5025);
            \addplot [color=blue_sw, line width=1pt, dashed, domain=-120:220,forget plot] (x,-1.015*x+20.176);
            \addplot [color=orange_sw, line width=1pt, dashed, domain=-120:220,forget plot] (x,-2.124*x+30.681);
            
            \addplot [color=gray!50, line width=1pt, dashed, domain=220:260] (x,-0.5114*x+6.5025);
            
            \legend{10 GW, 20 GW, Trend lines}
        \end{axis}
         % help grid
        % \draw[help lines,step=.2] (0,0) grid (16,4);
        % \draw[help lines,line width=.6pt,step=1] (0,0) grid (16,4);
        
        \node at (15.2,0.8) {\scriptsize \textcolor{gray!90}{FR}};
        \node at (9.4,0.4) {\scriptsize \textcolor{gray!90}{UK}};
        \node at (8.8,0.86) {\scriptsize \textcolor{gray!90}{BG}};
        \node at (8,0.8) {\scriptsize \textcolor{gray!90}{SK}};
        \node at (7.4,2.5) {\scriptsize \textcolor{gray!90}{NL}};
        \node at (4.9,2.7) {\scriptsize \textcolor{gray!90}{BE}};
        \node at (3.3,1.3) {\scriptsize \textcolor{gray!90}{IT}};
        \node at (1,2.7) {\scriptsize \textcolor{gray!90}{DE}};
        % \node at (1.3,1.9) {\scriptsize \textcolor{gray!90}{UK}};
        % \node at (1.4,3) {\scriptsize \textcolor{gray!90}{BE}};
        % \node at (2,2.4) {\scriptsize \textcolor{gray!90}{NL}};
        % \node at (2.9,2.3) {\scriptsize \textcolor{gray!90}{CH}};
        % \node at (2.4,3.1) {\scriptsize \textcolor{gray!90}{DE-LU}};
        % \node at (4.1,2) {\scriptsize \textcolor{gray!90}{AT}};
    \end{tikzpicture}
    }%
    % \vspace{-1.9em}
    \caption{Relation between import/export and social welfare variation for all price zones. Positive x coordinates refer to exports, while negative ones to imports.}
    \label{fig:sw}
    % \vspace{-1em}
\end{figure*}
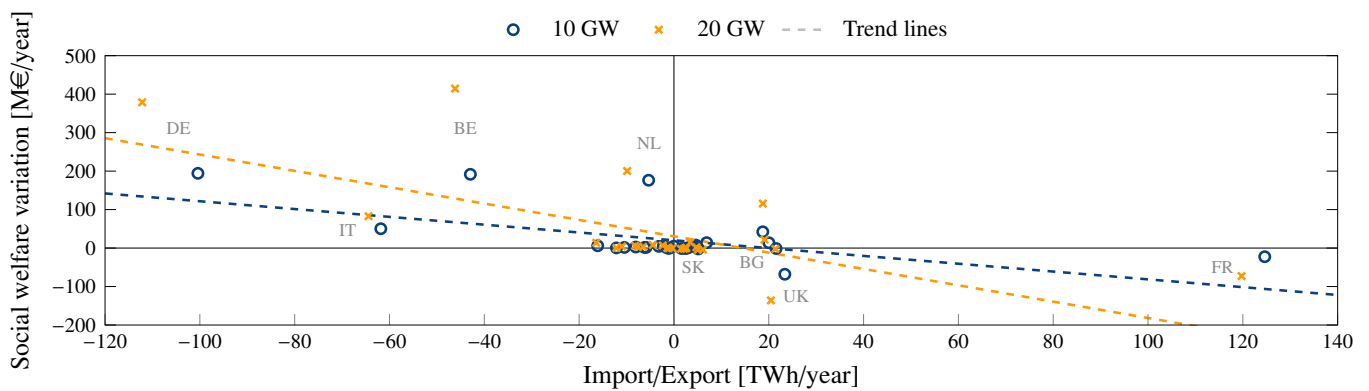

\subsection{Impact on the other European countries}

Once the impact on the countries directly connected to the hub has been investigated, the scope of the analysis is broadened to consider the rest of the European countries. Being an interconnected system, changes in the market equilibrium are to be expected also in areas far from the North Sea. 

The focus of this analysis is on market players, looking at consumer and producer surplus in the other countries. \figurename~\ref{fig:rev_OTHERS} shows the variation of producer (upper graph) and consumer (lower graph) surplus in the countries not directly connected to the hub across all the different scenarios. As it appears from \figurename~\ref{fig:rev_OTHERS}, variations occur in almost all countries. In particular, changes are more pronounced in those countries with high electricity prices, such as Austria, Switzerland and Italy. The most remarkable effect is, indeed, the gradual decrease of the congestions between Austria and Hungary. This happens because, by changing the market equilibrium, flows on all interconnectors change. For example, a certain portion of the flows from Poland and Czech Republic to Germany passes through the corridor between Austria and Hungary. With Germany decreasing the imports from these countries, the congestion is often relieved, resulting in lower prices in Austria, Switzerland, Italy and Slovenia and higher prices in Hungary. In the case of Belgium, instead, imports from the Netherlands increase significantly, reducing electricity prices. 

Other aspects to be noticed are: (i) electricity prices in Ireland are strongly correlated with prices in the UK and they follow a similar trend across the scenarios, (ii) electricity prices in Sweden and Finland have similar trends with prices in Norway and (iii) in the Balkan Peninsula generators surplus decrease outdoes consumer surplus increase. 

\definecolor{yellow_plot}{rgb}{1.0000,0.7529,0}%

\setcounter{figure}{11}

\begin{figure*}[!b]
\resizebox{0.98\textwidth}{!}{%
    \begin{tikzpicture}
        \begin{axis}[
            width  = 1\textwidth,
            height = 0.15\textheight,
            at={(1.20in,0.80601in)},
            scale only axis,
            axis on top,
            scaled y ticks = false,
            xmin=0,
            xmax=8760,
            xlabel style={font=\color{white!15!black}},
            xlabel={Hours (h)},
            xtick pos=left,
            ytick pos=left,
            ymin=0,
            ymax=1.2,
            every tick label/.append style={font=\footnotesize},
            ylabel style={font=\color{white!15!black},yshift=-5pt},
            ylabel={Normalized output},
            axis background/.style={fill=white},
            legend columns=3,
            legend style={at={(0.5,1.010001)}, anchor=south, legend cell align=left, align=left, draw=none, font=\small},
            /tikz/every even column/.append style={column sep=0.5cm},
            every axis legend/.append style={column sep=0.3em},
            clip mode=individual
            ]
            % \addplot[forget plot] coordinates {(0.60, 0.85)} node[above] {\textcolor{gray2}{\scriptsize{\textsc{overload}}}};
            \addplot[draw=blue_map, densely dashed, line width=1pt]table[row sep=crcr] {%
                0   2\\
                6   2\\
                };
            \addplot[draw=orange_plot, line width=1pt] table[y=$c$, x=$h$]{Plots/Data/cycling_20.dat};
            \addplot[draw=blue_map, densely dashed, line width=1pt, forget plot] table[y=$c$, x=$h$]{Plots/Data/cycling_nohub.dat};
            \legend{No Hub, 20 GW}
            % \addplot[draw=YellowFLOW, line width=0.8pt] table[y=$DE-1$, x=$t$]{Plots/Data/GFGforming_activepower.dat};
            % \addplot[draw=BlueFLOW, dashed, line width=0.8pt] table[y=$DE-2$, x=$t$]{Plots/Data/GFGforming_activepower.dat};
            % \addplot[draw=LBlueFLOW, line width=0.8pt] table[y=$DK$, x=$t$]{Plots/Data/GFGforming_activepower.dat};
            % \addplot[draw=GreenFLOW, line width=0.8pt] table[y=$NO$, x=$t$]{Plots/Data/GFGforming_activepower.dat};
            % \legend{Overload,NL,DE-1,DE-2,DK,NO}
        \end{axis}

    \end{tikzpicture}
    }%
    \vspace{-0.8em}
    \caption{Cycling pattern of a sample unit in NL.} 
    \label{fig:cycling}
    \vspace{-1em}
\end{figure*}  

Overall, social welfare increases in all four scenarios, respectively by 2027, 3027, 2544 and 2471 million Euros per year (between 0.01 and 0.03 percentage points) compared to the reference case. However, it must be considered that not in all countries the change is positive: as mentioned above, the UK experiences a decrease of social welfare in the first three scenarios. In general, countries that export a significant amount of energy (in relation to the internal consumption), e.g., France, the UK, and several others, experience decreases in producer surplus that are higher than the corresponding increases in consumer surplus. The relation between import/export and social welfare variation is depicted in \figurename~\ref{fig:sw}, where the negative part of the x-axis shows the import and the positive one the exports. The trend lines clearly show that social welfare tends to decrease with increasing exports. This happens in Bulgaria, Croatia, Montenegro, Slovenia and Slovakia in all scenarios, while in the UK and in France in some of the scenarios (mostly the first two).

\definecolor{orange_plot}{rgb}{0.964,0.572,0}
\definecolor{red_plot}{rgb}{0.874,0.325,0.152}
\definecolor{blue_plot}{rgb}{0.3020,0.7451,0.9333}%
\definecolor{green_plot}{rgb}{0.572,0.815,0.313}

\definecolor{blue_plot1}{rgb}{0.109,0.6784,0.8941}
\definecolor{blue_plot2}{rgb}{0.1490,0.5137,0.7765}
\definecolor{green_plot1}{rgb}{0.2588,0.7294,0.5922}
\definecolor{green_plot2}{rgb}{0.2431,0.5333,0.3255}

\definecolor{red_plot1}{rgb}{0.9098,0.2980,0.1333}%
\definecolor{orange_plot1}{rgb}{1.0000,0.7412,0.2784}%
\definecolor{red_plot2}{rgb}{0.7137,0.2863,0.1490}%
\definecolor{orange_plot2}{rgb}{1.0000,0.5176,0.1529}%

\setcounter{figure}{10}

\begin{figure}[!t]
    \centering
    \resizebox{0.48\textwidth}{!}{%
    \begin{tikzpicture}
    
    \begin{axis}[
            anchor = south west,
            width  = 0.49\textwidth,
            height = 0.200001\textheight,
            at={(0,0)},
            minor y tick num=3,
            ymajorgrids,
            yminorgrids,
            bar shift auto,
            ylabel = {Price [\euro/MWh]},
            ylabel style = {font=\color{black}},
            yticklabel style = {font=\color{black}\footnotesize},
            xticklabel style = {font=\color{black}\footnotesize},
            ytick style={black},
            y axis line  style={black,line width=1.0pt},
            xtick pos=left,
            xmin=0.5,xmax=4.5,
            xtick = {1,2,3,4},
            xticklabels={{10 GW},{20 GW},{Exch.},{UK}},
            ytick = {0,10,20,30,40,50},
            ymin=0,ymax=50,
            axis line style = {line width=0.1},
            legend columns=4,
            legend style={at={(0.5,1.01)}, anchor=south, legend cell align=left, align=left, draw=none, font=\small},
            every axis legend/.append style={column sep=0.3em},
            legend image code/.code={
                \draw [/tikz/.cd,bar width=5pt,yshift=-0.2em,bar shift=0pt]
                plot coordinates {(0cm,0.5em)};
            },
            ]
            \addplot[ybar, bar width=0.14, fill=blue_plot, draw=blue_plot, line width=0.05pt] table[row sep=crcr] {%
                1	36.07358658	\\
                2	25.37425931	\\
                3	43.07833836	\\
                4	28.2001975	\\
                };
            \addplot[ybar, bar width=0.14, fill=red_plot, draw=red_plot, line width=0.05pt] table[row sep=crcr] {%
                1	5.716672151	\\
                2	7.249706915	\\
                3	2.129055997	\\
                4	11.1229808	\\
                };
            \addplot[draw=black!50, forget plot, line width = 0.3] table[row sep=crcr] {%
                0   0\\
                5   0\\
                };
            \legend{Wind, HVDC};
        \end{axis}
\end{tikzpicture}
    }%
    % \vspace{-1.9em}
    \caption{Wind and HVDC capture prices at the hub across the different scenarios.}
    \label{fig:rev_NSEH}
    % \vspace{-1em}
\end{figure}        

\subsection{Revenues of wind producers and TSO on the island}

The focus is now shifted towards the revenues of the wind producers and a hypothetical system operator on the island. 

The capture prices of the offshore wind farms located at the NSEH, as well as the captured price difference of the HVDC lines connected to the hub, are shown in \figurename~\ref{fig:rev_NSEH}. As expected, the cannibalization effect of RES results in lower capture prices with more installed generation capacity, as electricity prices are likely to be lower during periods with high RES production. Indeed, in a future electricity market with only zero-marginal cost RES, other market mechanisms will be necessary for power producers to make profits, as there will be no more conventional generators to push prices up \citep{BARROSO2021,STRBAC2021}. Compared to the first scenario, the changes in the third and fourth scenarios can be explained in relation to the example provided in Section \ref{sec:2}. Indeed, in the third scenario additional transmission capacity is built in the North Sea, with the result that less congestions are created. Due to the assumption that wind producers bid at zero-marginal cost, the price in the island is formed based on the links that are not congested. Reducing congestions means that electricity prices converge to prices in the neighboring markets, resulting in higher prices at the hub and higher profits for wind producers. In the fourth scenario, instead, the connection to the UK leads to lower prices at the hub, as congestions occur more often on the links to the other countries. 

Opposite trends can be observed for the revenues of the system operator. Congestion rents are calculated as the energy exchanged on the links times the price difference between the connected market zones. The resulting revenues are then shared equally between the owners of the lines. Price differences arise only when congestions occur, meaning that the less congestions occur the lower the revenues of the transmission operators are. At the same time, the greater the price differences between market zones are, as in the case with the connection to the UK, the higher the revenues are. 

It is not clear yet whether there will be an independent system operator on the island, or this will be operated by the TSOs involved in the projects. Our analysis shows that the total revenues collected by all involved system operators are estimated between 220 million EUR/year and 920 million EUR/year. For example for the 10 GW island, the total revenues are 420 million EUR/year. Considering that each 1 GW line has a cost of about 1 billion Euros, a 10 GW island will require transmission investments of about 10 billion Euros. It becomes obvious that if we follow the merchant transmission investment model, the payback period will be over 20 years; such a payback period might make the merchant transmission model unattractive. On the other hand, we expect that the return rate regulation model will indeed be a commercially feasible option. As we show in our study in this paper, the construction of the energy island, including the necessary offshore transmission, is beneficial for the European social welfare and therefore will probably justify an investment based on the return rate regulation model. The revenues generated from the congestion rent can well be used to cover operation and maintenance costs of the offshore infrastructure or finance future expansion.

\subsection{Potential challenges arising with hybrid projects in the North Sea}

The analysis has focused so far on the direct market consequences of offshore hybrid projects like the NSEH. The analysis is now shifted towards challenges that might arise in terms of system and market operation. 

The first consideration concerns the flexibility of conventional generators and their cycling patterns. The high dependence of RES on weather parameters, such as wind speed, solar radiation, etc, makes them highly unpredictable and volatile. Sudden variations of power production challenge system operation if there are not enough sources of flexibility, e.g. energy storage or demand response. In our market model, technical limitations of generators, such as ramping limits and minimum online and offline duration, are not taken into consideration. This assumption could be acceptable for studying the impact on market players; however, it does not cover the feasibility aspect of the market outcome. Knowing the limitations of the model, we have performed this analysis ex-post, looking at the cycling patterns of several generators. The analysis has been carried out by counting the number of times each generator comes online and is shut down, i.e., that is considered a generation cycle, in the whole simulated year. \figurename~\ref{fig:cycling} shows the cycles of a biomass-fired power plant in the Netherlands. Compared to the reference case, cycles increase from 246 to 280 per year in the ``10 GW'' scenario, and to 307 in the ``20 GW'' scenario (respectively 14\% and 25\% higher). This increase is also experienced by other thermal units: the effect is more pronounced for generators on the left side of the merit order curve, namely nuclear, biomass and coal plants, which see the cycling increasing by 5\% to 35\%. On the contrary, gas and oil-fired units see a decrease of the cycling. This shows how expensive peaking technologies are gradually phased-out in a market driven fashion. Indeed, already today a lot of thermal generators are not profitable anymore, but are not allowed to shut down for security of supply reasons. 

With our model, it is not possible to state whether the increase in cycling is feasible; however, two possible scenarios lie ahead. If these changes are not feasible, either constraints will force different market outcomes, introducing less optimal equilibria, or more sources of flexibility will be necessary to deal with the higher intermittence of RES. On the other side, if this is actually feasible, additional operation and maintenance costs will be incurred by generating companies \citep{UMMELS2007, VANDENBERGH2015}, with the most probable outcome that bidding strategies will change to reflect these additional costs and electricity prices will further increase during low-wind-power events.

Finally, another aspect to take into consideration concerns the feasibility of installing tens of GW of wind power capacity in the North Sea. The NSWPH consortium is studying the construction of several artificial islands (11 to 17) in the North Sea; the amount of wind power collected in each island ranges between 3 and 14 GW \citep{NSWPHCostEvaluation}. Some onshore connection points have been identified in \citet{CONNECTIONPOINTS}; however, in these analyses conducted it is simply assumed that these locations can accommodate such large amount of wind energy. Given that not all the points of connection are specified in the technical reports, the closest points in the geographical areas identified by \citet{CONNECTIONPOINTS} have been used in our transmission grid model. The projection of the market outcome obtained in the ``20 GW'' scenario has been performed as illustrated in Section \ref{sec:3_4}, with particular attention to the wind curtailment in the island. \figurename~\ref{fig:feas_wC} shows the amount of energy produced vs. energy curtailed. From our results, it seems there might be difficulties in integrating a large amount of wind energy in the North Sea unless grid reinforcements are performed and energy storage or Power-to-X (PtX) solutions are deployed. In particular, congestions arise in the German national grid, resulting in more than 2 TWh of curtailed wind energy. On the one hand, this analysis is highly impacted by the selected connection points; on the other hand, we are considering only two islands with 20 GW of installed wind power. Looking ahead at the next 30 years, more and more offshore wind capacity will be installed, calling for grid reinforcements, energy storage or PtX solutions.

\definecolor{blue_wc}{rgb}{0.3020,0.7451,0.9333}%
\definecolor{blue_wc2}{rgb}{0.0353,0.5373,0.6941}%
\definecolor{red_wc}{rgb}{0.75,0,0}%

\setcounter{figure}{12}

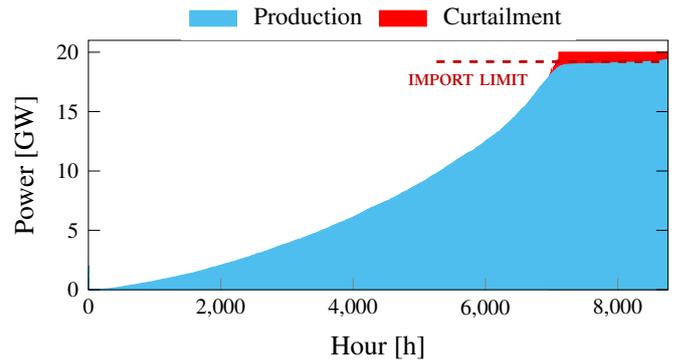
\begin{figure}[!t]
    \centering
    \resizebox{0.48\textwidth}{!}{%
    \begin{tikzpicture}
        \begin{axis}[
            width  = 0.49\textwidth,
            height = 0.20\textheight,
            at={(1.20in,0.806in)},
            axis on top,
            % axis y line*=left,
            % axis x line=box,
            % ybar=0.5pt,
            % bar width=0.25,
            % scale only axis,
            xlabel = {Hour [h]},
            ylabel = {Power [GW]},
            ylabel style = {font=\color{black}},
            yticklabel style = {font=\color{black}\footnotesize},
            xticklabel style = {font=\color{black}\footnotesize},
            ytick style={black},
            y axis line  style={black,line width=1.0pt},
            xtick pos=left,
            xmin=0,xmax=8760,
            xtick = {0,2000,4000,6000,8000},
            ytick = {0,5,10,15,20},
            ymin=0,ymax=21,
            axis line style = {line width=0.1},
            legend columns=2,
            legend style={at={(0.5,1)}, anchor=south, legend cell align=left, align=left, draw=none, font=\small},
            every axis legend/.append style={column sep=0.4em},
            legend image code/.code={
                \draw [/tikz/.cd,bar width=5pt,yshift=-0.2em,bar shift=0pt]
                plot coordinates {(0cm,0.5em)};
            },
            ]
                \addplot[area legend, draw=blue_wc, fill=blue_wc]table[row sep=crcr] {%
                0   2\\
                6   2\\
                }\closedcycle;
            \addplot[area legend, draw=red, fill=red] table[y=$w_c$, x=$h$]{Plots/Data/winC.dat}\closedcycle;
            \addplot[area legend, draw=blue_wc, fill=blue_wc, forget plot] table[y=$w_p$, x=$h$]{Plots/Data/winP.dat}\closedcycle;
            \addplot[draw=red_wc, line width = 1, dashed, forget plot]table[row sep=crcr] {%
                5260  19.2\\
                8760  19.2\\
                };
            \legend{Production, Curtailment}
        \end{axis}
        \node[] at (7.9,4.75) {\small \textcolor{red_wc}{\textsc{import limit}}};
    \end{tikzpicture}
    }%
    % \vspace{-1.9em}
    \caption{Duration curve of produced and curtailed wind power in the ``20~GW'' scenario. The red dashed line corresponds to the technical limitation imposed by congestions in national grids.}
    \label{fig:feas_wC}
    % \vspace{-1em}
\end{figure}

% \vspace{-0.5em}
\section{Conclusion and policy implications}\label{sec:5}
In this paper, we have presented detailed market analyses related to the realization of hybrid offshore projects in the North Sea, inspired by the North Sea Wind Power Hub (NSWPH) programme. Given the large number of countries interconnected through the pan-European AC transmission grid, there is the need for large-scale impact analyses of these projects on the European power system.

To this end, two detailed transmission grid and electricity market models representing the European power system were developed and presented in this paper. Four different scenarios were designed to investigate different connection topologies and hub sizes. The results show that, overall, social welfare increases in all the considered scenarios. Social welfare here has been expressed with focus on the operational stage during the shorter term of a year, which can then be compared with the cost of infrastructure investments; for this a longer lifetime of several decades must be considered. At the same time, we have presented evidence of several challenges that might arise in terms of system and market operation. Addressing those will allow for the efficient integration of North Sea Energy Islands, and lead to an increase in social welfare.

From a market point of view, we showed that not all countries register a positive increase of welfare and we pointed at the fact that the change in social welfare were heterogeneous among countries over the European area considered. This stems from the consideration that exporting countries experience (i) lower electricity prices and/or (ii) lower exports due to the introduction of the energy island. Therefore, the increase in consumer surplus due to lower prices is not enough to balance out the decrease of producer surplus due to lower exports and prices. Given that benefits shift from producers to consumers, it may be that the return on investment may not be sufficient on the producer side. Specific mechanisms may need to be thought of to better support investment on the generation side then, e.g., through capacity auction mechanisms (which have increased in popularity over the years), or by opening the possibility for them to invest in the construction of North Sea wind farms to maintain a revenue stream. Other approaches to support revenue streams which could be considered, but maybe less likely, would relate to crowd-funding systems in line with the recent development in peer-to-peer energy markets. In parallel, the heterogeneity in changes in social welfare may lead to disparities among European countries and the perceived benefits from investment in these new infrastructures. There, different form of local energy conversion and storage may allow to smoothen these differences by adding a temporal dimension to arbitrage opportunities. In this context, Power-to-X (PtX) is probably the most promising technology for bulk storage.

On a system perspective, we showed that conventional generators and national grids are stressed by the large amount of wind capacity installed in the North Sea. Conventional generators experience increasing cycling as more and more intermittent RES generation is integrated. This could lead to significant changes in the bidding strategies of market players to reflect the additional O\&M costs, or might require additional sources of flexibility. Therefore, there is the need of intensified investments in flexible generation or in upgrades of existing generators that can increase their flexibility or the number of cycles. Moreover, because of internal congestions, it might not be feasible to import large amounts of wind energy unless grid reinforcements are performed. In this context, again, incentives for energy storage solutions or other sources of flexibility, e.g. PtX used to absorb excess wind energy \cite{Singlitico_2021}, could help relieve some congestions and increase the exploitation of wind resources in the North Sea.
% \vspace{-0.5em}

\appendix
\section{PTDF matrix for flow-based market coupling}\label{sec:appendix1}
Power Transfer Distribution Factors (PTDFs) are sensitivity coefficients that show how power flows are distributed following the injection of one unit of power at a specific bus (the source) and its withdrawal at the reference bus (the sink). Under the assumption of ``DC'' power flow approximation, power flows $f$ across the network can be calculated as follows:
\begin{equation}\label{eq:flows}
    f = \text{PTDF}\,J
\end{equation}
where $J$ is the vector of power injections and $\rm{PTDF}$ is the matrix of power transfer distribution factors. The PTDF matrix is an $L\times N$ matrix, with $N$ the number of buses and $L$ the number of lines.

The PTDF matrix of a system depends on the network connectivity and on the line parameters. The change in the flow over the line connecting bus $i$ to bus $j$ due to a unit injection at bus $n$ is given by:
\begin{equation}\label{eq:PTDF}
    \rm{PTDF}_{ij,n} = \frac{X_{i,n}-X_{j,n}-X_{i,r}+X_{j,r}}{x_{i,j}}
\end{equation}
with $r$ indicating the reference node, $x_{i,j}$ the reactance of the transmission line connecting bus $i$ to bus $j$ and $X_{i,k}$ the entry in the $i$th row and the $k$th column of the bus reactance matrix $\boldsymbol{X}$.

When the system is reduced to represent a zonal market, the definition of PTDF in \eqref{eq:PTDF} is not valid anymore. The procedure described in Algorithm~\ref{alg:PTDF} is followed to estimate the new $\text{PTDF}^{\,*}$ matrix of the system. One at a time, the output $P_g$ of all generators in one bidding zone (the source) is increased by one unit of power, e.g. 1 MW, and the flows on all the lines are calculated using \eqref{eq:flows}, where the PTDF matrix is the one corresponding to the full network model. The flows on the set of lines corresponding to the interconnector between two bidding zones are aggregated and saved. The same operation is repeated varying the bus $n$ at which that unit of power is consumed ($D_n$), until all the buses within the reference bidding zone (the sink) have been considered (from 1 to $N_r$). At the end, the column with the PTDFs corresponding to the considered bidding zone (the source) is estimated by statistical analysis using linear regression. The whole procedure is repeated for another bidding zone (corresponding to another column of the matrix), until all the bidding zones have been considered (from 1 to $Z$). The resulting $\text{PTDF}^{\,*}$ matrix is an $I\times Z$ matrix, with $Z$ the number of bidding zones and $I$ the number of (equivalent) interconnectors.

With this procedure, the estimated $\text{PTDF}^{\,*}$ matrix should provide a good enough representation of how the flows on the interconnectors change when the net position of a bidding zone change. Thus, a flow-based market coupling algorithm is obtained by calculating the power exchanges between bidding zones with the new $\text{PTDF}^{\,*}$ matrix.

\begin{algorithm}[!t]
\caption{PTDF matrix estimation}\label{alg:PTDF}
\begin{algorithmic}[1]
\FOR{$z \gets 1$ \TO $Z$} 
\FOR{$g \gets 1$ \TO $G_z$} 
\STATE Set $P_g$ = 1
\FOR{$n \gets 1$ \TO $N_r$} 
\STATE Set $D_n$ = 1
\STATE Solve power flow
\STATE Aggregate flows on interconnectors
\STATE Save flows on interconnectors
\ENDFOR
\ENDFOR
\FOR{$int \gets 1$ \TO $I$} 
\STATE Calculate $\text{PTDF}^{\,*}_{int,z} = \sum f_{int} / G_zN_r$
\ENDFOR
\ENDFOR
\end{algorithmic}
\end{algorithm}

Since the procedure involves statistical analysis using linear regression, the obtained $\text{PTDF}^{\,*}$ might contain small numerical inaccuracies. If this happens, the power balance constraint that is implicitly included in the market clearing problem does not hold. In order to solve this issue, the following linear program can be used:

\begin{subequations}\label{eq:3_newopt}
    \begin{alignat}{3}
        \underset{w,z}{\text{min}} \enspace 
        & \sum_{int}\sum_{z}w_{int,z}^2 + \sum_{int}\sum_{k}\pi y_{int,k}^2 \label{PTDF:obj} \\
        \text{s.t.} \enspace & |w_{int,z}| \leq 0.1 && \quad :\,\,\forall int,\,\forall z \label{PTDF:lim1}\\
        & |y_{int,k}| \leq 0.1 && \quad :\,\,\forall int,\,\forall k \label{PTDF:lim2}\\
        & J_k - I^{int}({\rm PTDF^{\,*}}+ w)J_k = 0 && \quad :\,\,\forall k \label{PTDF:balance}\\
        & ({\rm PTDF^{\,*}}+ w)J_k - {\rm PTDF^{\,*}}J_k = y_k && \quad :\,\,\forall k  \label{PTDF:flows}
    \end{alignat}
\end{subequations}
where $w_{int,z}$ is the correction of the $\text{PTDF}^{\,*}$ element corresponding to interconnector $int$ and bidding zone $z$, $y_{int,k}$ is the tolerance on the flow precision, $\pi$ are weights and $I^{int}$ is the incidence matrix of the interconnectors on the bidding zones. The vector $J_k$ contains the net position of each bidding zone (i.e. the power injections), and the index $k$ correspond to a specific set of injections (in our case $k\in\{1,...,1000\}$, each time with randomized injections). The optimization problem aims at minimizing the deviation from the $\text{PTDF}^{\,*}$ while keeping the error on the flows relatively small. Constraints \eqref{PTDF:balance} enforce the power balance for all the sets of power injections, while constraints \eqref{PTDF:flows} enforce the equality on the flows calculated with the ``old'' and ``new'' $\text{PTDF}^{\,*}$.

The final ${\rm PTDF^{\,f}}$ matrix for flow-based market coupling is calculated as:
\begin{equation}
    {\rm PTDF^{\,f}} = \text{PTDF}^{\,*} + w
\end{equation}

\section{Mathematical formulation}\label{sec:appendix2}

For replicability of the results, the optimization problems used to clear the day-ahead market and project the dispatch on the full network model are presented here. 

The market clearing problem is formulated as the following linear program:
\begin{subequations}\label{eq:MCP}
    \begin{alignat}{3}
        \underset{g,h,d,d^{\rm s},w^{\rm c},f^{\rm dc}}{\text{max}} & \quad u^\intercal d - c^{\rm g \,\intercal} g - c^{\rm h \,\intercal} h - v^{\rm d \,\intercal} d^{\rm s} && - v^{\rm w \,\intercal} w^{\rm c} \label{MCP:obj} \\
        \text{s.t.} & \quad 0 \leq g \leq \overline{G} && \label{MCP:gen} \\
        & \quad \underline{H}\color{black} \leq h \leq \overline{H} && \label{MCP:hydro} \\
        & \quad 0 \leq re^{\rm c} \leq W+S+R && \label{MCP:windc} \\
        & \quad 0 \leq d \leq 0.2D && \label{MCP:eldem} \\
        & \quad 0 \leq d^s \leq D && \label{MCP:loadshed} \\
        & \quad res = RES^{\rm t-1}\, + INF - \text{I}^{\rm h} h && \label{MCP:water_balance} \\
        & \quad RES^{\rm trj} \leq res \leq \overline{RES} && \label{MCP:min_res} \\
        & \quad f^{\rm ac} = \text{PTDF}^{\rm f}\,J && \quad : \, \varphi \label{MCP:flowAC} \\
        & \quad \underline{F}^{\rm ac} \leq f^{\rm ac} \leq \overline{F}^{\rm ac} && \label{MCP:fac} \\
        & \quad \underline{F}^{\rm dc} \leq f^{\rm dc} \leq \overline{F}^{\rm dc} && \label{MCP:fdc} \\
        & \quad \textstyle\sum_{z=1}^{Z} J_z = 0 && \quad : \, \lambda \label{MCP:balance}
    \end{alignat}
\end{subequations} %
where the optimization variables are generator outputs $g$, hydro unit outputs $h$, water reservoir levels $res$\color{black}, elastic demand consumption $d$, load shedding values $d^{\rm s}$, RES curtailment values $re^{\rm c}$ and HVDC set-points $f^{\rm dc}$. The input parameters are wind, solar and hydro run-of-river \color{black} power outputs, respectively $W$, $S$ and $R$\color{black}, inelastic consumption levels $D$, AC and DC transmission capacities in the two directions, respectively $\underline{F}^{\rm ac}$, $\overline{F}^{\rm ac}$ and $\underline{F}^{\rm dc}$, $\overline{F}^{\rm dc}$, water reservoir level at time ($t$-1) $RES^{\rm t-1}$, lower and upper bounds of the water reservoirs, respectively $RES^{\rm trj}$ and $\overline{RES}$, water inflow into the reservoirs $INF$, minimum and maximum hydro generation levels, respectively $\underline{H}$ and $\overline{H}$,\color{black} generator capacities $\overline{G}$ and the PTDF matrix for flow-based market coupling ${\rm PTDF}^{\rm f}$. The objective function \eqref{MCP:obj} is the difference between load utilities, with $u$ the linear utility coefficients, and the sum of production costs, with $c^{\rm g}$ and $c^{\rm h}$ the linear cost coefficients of conventional generators and hydro units, and the costs associated with load shedding and wind curtailment, indicated by $v^{\rm s}$ and $v^{\rm w}$ respectively. Constraints \eqref{MCP:gen}-\eqref{MCP:loadshed} and \eqref{MCP:fdc} enforce the lower and upper bounds on optimization variables, while Constraints \eqref{MCP:water_balance} represent water balance constraints. For pumped-hydro storage plants, the lower bound of generation is negative; when water is pumped into the reservoir, $h$ is negative and the reservoir level increases. Constrains \eqref{MCP:min_res} enforce lower bounds on the reservoir levels, making sure reservoir trajectories are followed (see \figurename~\ref{fig:reservoir}). \color{black} Constraints \eqref{MCP:flowAC} define the flows on AC interconnectors as the product of the PTDF matrix and the vector $J$ containing the net positions of the bidding zones, calculated as

\definecolor{orange_plot}{rgb}{0.964,0.572,0}
\definecolor{red_plot}{rgb}{0.874,0.325,0.152}
\definecolor{blue_plot}{rgb}{0,0.439,0.752}
\definecolor{green_plot}{rgb}{0.572,0.815,0.313}

\definecolor{blue_plot1}{rgb}{0.109,0.6784,0.8941}
\definecolor{blue_plot2}{rgb}{0.1490,0.5137,0.7765}
\definecolor{green_plot1}{rgb}{0.2588,0.7294,0.5922}
\definecolor{green_plot2}{rgb}{0.2431,0.5333,0.3255}

\definecolor{red_plot1}{rgb}{0.9098,0.2980,0.1333}%
\definecolor{orange_plot1}{rgb}{1.0000,0.7412,0.2784}%
\definecolor{red_plot2}{rgb}{0.7137,0.2863,0.1490}%
\definecolor{orange_plot2}{rgb}{1.0000,0.5176,0.1529}%

\definecolor{yellow_plot1}{rgb}{1.0000,0.9176,0.6118}%
\definecolor{brown_plot1}{rgb}{0.8078,0.5529,0.2431}%
\definecolor{yellow_plot2}{rgb}{1.0000,0.7922,0.0314}%
\definecolor{brown_plot2}{rgb}{0.5176,0.4039,0}%

\definecolor{green2_plot}{rgb}{0.329411764705882,0.619607843137255,0.223529411764706}%
\definecolor{green3_plot}{rgb}{0.541176470588235,0.721568627450980,0.200000000000000}%
\definecolor{green1_plot}{rgb}{0.752941176470588,0.811764705882353,0.227450980392157}%
\definecolor{green4_plot}{rgb}{0.00784313725490196,0.588235294117647,0.462745098039216}%

\begin{figure}[!t]
    \centering
    \resizebox{0.48\textwidth}{!}{%
    \begin{tikzpicture}
        \begin{axis}[
            anchor = north,
            width  = 0.49\textwidth,
            height = 0.200001\textheight,
            at={(0,0)},
            ylabel = {Energy index},
            xlabel = {Week},
            ylabel style = {font=\color{black}},
            yticklabel style = {font=\color{black}\footnotesize},
            xticklabel style = {font=\color{black}\footnotesize},
            ytick style={black},
            y axis line  style={black,line width=1},
            xtick pos=left,
            xmin=0,xmax=53,
            ymin=0.4,ymax=0.8,
            axis line style = {line width=0.1},
            legend columns=4,
            legend style={at={(0.5,1.01)}, anchor=south, legend cell align=left, align=left, draw=none, font=\footnotesize},
            every axis legend/.append style={column sep=0.3em},
            legend image code/.code={
                \draw[mark repeat=2,mark phase=2]
                plot coordinates {
                    (0cm,0cm)
                    (0.2cm,0cm)        %% default is (0.3cm,0cm)
                    (0.4cm,0cm)         %% default is (0.6cm,0cm)
                    };
                }
            ]
            \addplot[draw=black!50, forget plot, line width = 0.3] table[row sep=crcr] {%
                0   0\\
                53   0\\
                };
            \addplot[draw=green4_plot, line width=1.3] table[row sep=crcr] {%
                0	0.540977509	\\
                1	0.56682627	\\
                2	0.580565788	\\
                3	0.568084111	\\
                4	0.507053217	\\
                5	0.482369299	\\
                6	0.498013593	\\
                7	0.517025572	\\
                8	0.516693655	\\
                9	0.539290536	\\
                10	0.574001129	\\
                11	0.590933097	\\
                12	0.593776809	\\
                13	0.588407799	\\
                14	0.588080842	\\
                15	0.595066396	\\
                16	0.646331304	\\
                17	0.69668772	\\
                18	0.686632614	\\
                19	0.725022689	\\
                20	0.723406004	\\
                21	0.675452245	\\
                22	0.645457892	\\
                23	0.666770672	\\
                24	0.685351084	\\
                25	0.666466739	\\
                26	0.626321553	\\
                27	0.622244812	\\
                28	0.63378676	\\
                29	0.629391914	\\
                30	0.625005418	\\
                31	0.613543335	\\
                32	0.597871284	\\
                33	0.58549831	\\
                34	0.572771786	\\
                35	0.561588617	\\
                36	0.53841499	\\
                37	0.519123467	\\
                38	0.555996331	\\
                39	0.58026518	\\
                40	0.545916897	\\
                41	0.508804398	\\
                42	0.507542641	\\
                43	0.526455813	\\
                44	0.505657346	\\
                45	0.498396581	\\
                46	0.490935105	\\
                47	0.493617699	\\
                48	0.491471352	\\
                49	0.496870958	\\
                50	0.485356593	\\
                51	0.486536469	\\
                52	0.539472299	\\
                };
            \addplot[draw=green1_plot, line width=1.3] table[row sep=crcr] {%
                0	0.540977509	\\
                1	0.56682627	\\
                2	0.580568917	\\
                3	0.568130167	\\
                4	0.50788527	\\
                5	0.49198866	\\
                6	0.511054618	\\
                7	0.536603209	\\
                8	0.523796455	\\
                9	0.560612181	\\
                10	0.603333358	\\
                11	0.640241152	\\
                12	0.607618265	\\
                13	0.601204923	\\
                14	0.62321378	\\
                15	0.610879461	\\
                16	0.672639349	\\
                17	0.741328258	\\
                18	0.695399654	\\
                19	0.742311508	\\
                20	0.741429707	\\
                21	0.695459024	\\
                22	0.667578753	\\
                23	0.666782978	\\
                24	0.698631142	\\
                25	0.666466739	\\
                26	0.634194406	\\
                27	0.622263394	\\
                28	0.646474972	\\
                29	0.631025592	\\
                30	0.631807094	\\
                31	0.622122954	\\
                32	0.599106399	\\
                33	0.58549831	\\
                34	0.572771786	\\
                35	0.580424148	\\
                36	0.54034494	\\
                37	0.520155663	\\
                38	0.556007582	\\
                39	0.58222973	\\
                40	0.560708656	\\
                41	0.509665412	\\
                42	0.511748499	\\
                43	0.533207327	\\
                44	0.505928155	\\
                45	0.498396581	\\
                46	0.490935105	\\
                47	0.496797438	\\
                48	0.500478655	\\
                49	0.509831118	\\
                50	0.490599499	\\
                51	0.490501484	\\
                52	0.541026482	\\
                };
            \legend{Minimum level, Output level}
        \end{axis}
    \end{tikzpicture}}
    % \vspace{-0.5cm}
    \caption{Example of minimum reservoir level and output level of the model.}
    \label{fig:reservoir}
    % \vspace{-1em}
\end{figure}
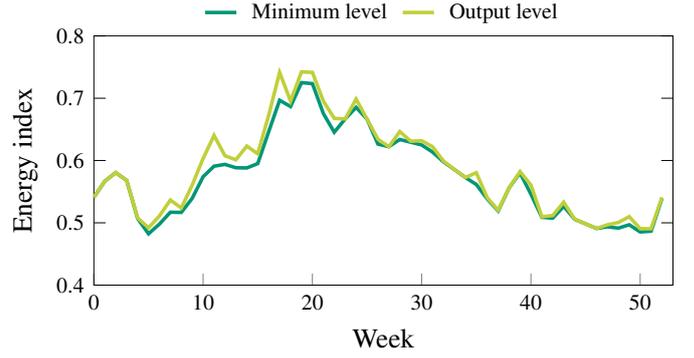

\begin{equation}
    J = \text{I}^{\rm g} g + \text{I}^{\rm h} h + RE \,- \text{I}^{\rm d}(D + d - d^{\rm s}) - \text{I}^{\rm dc} f^{\rm dc}
\end{equation}
where $\text{I}^{\rm g}$ and $\text{I}^{\rm h}$ are the incidence matrices of thermal and hydropower generators on the bidding zones, $\text{I}^{\rm d}$ is the incidence matrix of loads and $\text{I}^{\rm dc}$ is the incidence matrix of HVDC interconnectors. $RE$ is the sum of the renewable generation, calculated as:
\begin{equation}
    RE = \text{I}^{\rm pv} S + \text{I}^{\rm w} W + \text{I}^{\rm r} R - re^{\rm c}
\end{equation}
where $\text{I}^{\rm pv}$, $\text{I}^{\rm w}$ and $\text{I}^{\rm r}$ are the incidence matrices of solar PV power stations, wind farms and run-of-river hydropower plants. \color{black} Constraints \eqref{MCP:fac} enforce the lower and upper bounds on power flows over AC interconnectors. Finally, constraint \eqref{MCP:balance} represents the power balance of the system. 
The dual variables associated with constraints \eqref{MCP:balance} and \eqref{MCP:flowAC} are used to compute the Locational Marginal Prices (LMP) for each bidding zone as follows:
\begin{equation}
    {\rm LMP} = \lambda + \text{PTDF}^{\rm f\,\intercal} \varphi
\end{equation}

Once the market is cleared, the set-points of generators and loads are defined as

\begin{align}
    G^* = & \, g^* \\
    H^* = & \, h^* \\
    RE^* = & \, \text{I}^{\rm w} W+\text{I}^{\rm pv} S + \text{I}^{\rm r} R - re^{\rm c\,*} \\
    D^* = & \, D+d^*-d^{\rm s\,*} \\
    RES^* = & \,res^*
\end{align}
\color{black}
The feasibility of the resulting dispatch is then checked by projecting the set-points on the transmission network model. The following linear program is used to calculate the necessary redispatching:
\begin{subequations}\label{eq:PP}
    \begin{align}
        \underset{g^{\pm},h^{\pm}, d^{\rm s}, w^{\rm c}, f^{\rm dc}}{\text{min}} &
        \quad c^{\rm g+ \,\intercal} g^+ + c^{\rm g- \,\intercal} g^- + c^{\rm h+ \,\intercal} h^+ \,+ \nonumber\\
        & \quad\quad + c^{\rm h- \,\intercal} h^- + v^{\rm d \,\intercal} d^{\rm s} + v^{\rm w \,\intercal} w^{\rm c} \label{PP:obj} \\
        \text{s.t.} & \quad 0 \leq g^+ \leq G^* \label{PP:gen+} \\
        & \quad 0 \leq g^- \leq \overline{G}-G^* \label{PP:gen-} \\
        & \quad 0 \leq h^+ \leq H^* \label{PP:hydro+} \\
        & \quad 0 \leq h^- \leq \overline{H}-H^* \label{PP:hydro-} \\
        & \quad 0 \leq re^{\rm c} \leq RE^* \label{PP:windc} \\
        & \quad 0 \leq d^s \leq D^* \label{PP:loadshed} \\
        & \quad res = RES^* - \text{I}^{\rm h} (h^+-h^-) && \label{MCP:water_balance} \\
        & \quad RES^{\rm trj} \leq res \leq \overline{RES} && \label{MCP:min_res} \\
        & \quad f^{\rm ac} = \text{PTDF}\,J \label{PP:flowAC} \\
        & \quad \underline{F}^{\rm ac} \leq f^{\rm ac} \leq \overline{F}^{\rm ac} \label{PP:fac} \\
        & \quad \underline{F}^{\rm dc} \leq f^{\rm dc} \leq \overline{F}^{\rm dc} \label{PP:fdc} \\
        & \quad \textstyle\sum_{n=1}^{N} J_n = 0 \label{PP:balance}
    \end{align}
\end{subequations} %
where the optimization variables are the up- and down-regulation of conventional generators, respectively $g^+$ and $g^-$, the up- and down-regulation of hydro power units, respectively $h^+$ and $h^-$, load shedding and wind curtailment values, respectively $d^{\rm s}$ and $re^{\rm c}$, and the flows on HVDC lines $f^{\rm dc}$. The redispatching costs are calculated as the deviation from the day-ahead price as follows:
\begin{equation}
    \begin{cases}
    c_i^{g+} = 0,\,\,c_i^{g-}=\text{LMP}_{z:i\in\mathcal{G}_z}-c_i, & \quad \text{if}\,\,g_i^* = \overline{G}_i \\
    c_i^{g+} = c_i-\text{LMP}_{z:i\in\mathcal{G}_z},\,\,c_i^{g-}=0, & \quad \text{if}\,\, g_i^* = 0 \\
    c_i^{g+} = 0, \,\,c_i^{g-}=0, & \quad \text{if}\,\, 0 \leq g_i^* \leq \overline{G}_i \\
    \end{cases}
\end{equation}
where the subscript $i$ refers to the $i$th generator and the subscript $z:i\in\mathcal{G}_z$ to the bidding zone $z$ where the generator $i$ is located. Redispatching costs for hydro power units are calculated in the same way. 

Note that constraints \eqref{PP:flowAC} calculate the flows on all the AC lines, with PTDF the full matrix calculated with Eq. \eqref{eq:PTDF} and $J$ the nodal injections, calculated as

\begin{equation}
\begin{split}
    J = \,& \text{I}^{\rm g} (G^*+g^+-g^-) + \text{I}^{\rm h} (H^*+h^+-h^-) \,+  \\ 
    & \quad  + RE^* +re^{\rm c}-\text{I}^{\rm d}(D^* - d^{\rm s}) - \text{I}^{\rm dc} f^{\rm dc}
\end{split}
\end{equation}
\color{black}
Finally, constraint \eqref{PP:balance} enforces the system balance.

\section{Numerical results}\label{sec:appendix3}

In this section, the numerical results are presented in differnet tables. This has a double purpose: provide the interested reader with additional information and support the replicability of results through validation. 

Table~\ref{tabA1:prices} shows the average electricity price per country across the five scenarios. Table~\ref{tabA2:gen} and \ref{tabA3:demand} show respectively the total amount of energy generated and consumed by each country. Finally, Table~\ref{tabA4:export} and \ref{tabA5:import} show the exports and the imports of each country.

\section*{Acknowledgement}
This work is partly supported by Innovation Fund Denmark through the multiDC project (Grant Agreement No. \mbox{6154-00020B}), by the EU project Best Paths under the 7th Framework Programme (Grant Agreement No. 612748), by the ForskEL-projekt 12264 Best Paths for DK, and by the EUDP project North Sea Energy Island Feasibility Study (Grant Agreement No. \mbox{64018-0584}).

Part of the European transmission model used in this paper was developed thanks to the efforts of Lejla Halilbasic, Florian Thams, Riccardo Zanetti and Georgia Tsoumpa within the framework of the European project Best Paths, and by one project partner, CIRCE.

A special thanks to Fabio Moret for his help with the visualization of the European grid.

%% BIBLIOGRAPHY 
 \bibliographystyle{unsrtnat} 
 \bibliography{00_main.bib}
 
\begin{table*}[!t]
    \caption{\normalsize \textbf{Average electricity prices} across the different scenarios (\textbf{Euros/MWh}).}
    \label{tabA1:prices}
    \centering
    \vspace{0.5em}
    \resizebox{!}{0.45\textwidth}{
    \begin{tabularx}{0.70\textwidth}{*1{>{\arraybackslash}X} *5{>{\centering\arraybackslash}X}}
        \hline
                  & {\bf No Hub} & {\bf 10 GW} & {\bf 20 GW} & {\bf Exchanges} & {\bf UK} \Tstrut\Bstrut\\ 
        \hline          
        AL	&	26.93	&	27.01	&	27.03	&	26.98	&	26.98	\Tstrut\\
        AT	&	49.25	&	47.18	&	45.11	&	46.93	&	47.04	\\
        BA	&	46.75	&	46.20	&	45.70	&	46.10	&	46.16	\\
        BE	&	49.22	&	45.36	&	40.75	&	45.34	&	44.97	\\
        BG	&	48.21	&	48.13	&	48.06	&	48.12	&	48.13	\\
        CH	&	55.63	&	53.88	&	52.80	&	53.27	&	53.88	\\
        CZ	&	54.01	&	53.06	&	52.23	&	52.95	&	52.99	\\
        DE-LU	&	48.83	&	46.33	&	43.43	&	45.97	&	46.17	\\
        DK	&	41.12	&	40.43	&	37.47	&	41.84	&	39.82	\\
        ES	&	35.36	&	35.15	&	34.93	&	35.18	&	35.16	\\
        FI	&	28.82	&	28.59	&	28.24	&	28.65	&	28.56	\\
        FR	&	25.99	&	25.68	&	25.30	&	25.80	&	25.83	\\
        GR	&	59.03	&	58.99	&	58.99	&	58.99	&	59.00	\\
        HR	&	50.30	&	49.52	&	48.73	&	49.40	&	49.47	\\
        HU	&	48.99	&	48.53	&	48.10	&	48.45	&	48.52	\\
        IE	&	15.36	&	14.94	&	14.26	&	15.12	&	19.54	\\
        IT	&	52.76	&	52.31	&	51.95	&	52.23	&	52.27	\\
        ME	&	44.74	&	44.53	&	44.34	&	44.51	&	44.53	\\
        MK	&	47.20	&	47.12	&	47.02	&	47.13	&	47.12	\\
        NL	&	43.42	&	41.92	&	36.18	&	43.93	&	41.82	\\
        NO	&	33.79	&	33.18	&	32.16	&	33.48	&	33.22	\\
        PL	&	58.67	&	58.45	&	58.22	&	58.42	&	58.42	\\
        PT	&	35.36	&	35.18	&	34.96	&	35.21	&	35.19	\\
        RO	&	48.44	&	48.23	&	48.02	&	48.19	&	48.22	\\
        RS	&	46.65	&	46.51	&	46.37	&	46.50	&	46.51	\\
        SE	&	29.97	&	29.51	&	28.92	&	29.61	&	29.45	\\
        SI	&	51.04	&	50.12	&	49.16	&	50.01	&	50.06	\\
        SK	&	49.65	&	49.17	&	48.75	&	49.09	&	49.14	\\
        UK	&	25.24	&	24.22	&	22.80	&	24.58	&	28.87	\\
        NSEH	&	0.00	&	40.14	&	34.60	&	43.80	&	35.84	\Bstrut\\
        \hline
   \end{tabularx}}
\end{table*}

\begin{table*}[!t]
    \caption{\normalsize \textbf{Total generation} across the different scenarios (\textbf{TWh/year}).}
    \label{tabA2:gen}
    \centering
    \vspace{0.5em}
    \resizebox{!}{0.45\textwidth}{
    \begin{tabularx}{0.70\textwidth}{*1{>{\arraybackslash}X} *5{>{\centering\arraybackslash}X}}
        \hline
                  & {\bf No Hub} & {\bf 10 GW} & {\bf 20 GW} & {\bf Exchanges} & {\bf UK} \Tstrut\Bstrut\\ 
        \hline          
        AL	&	13.45	&	13.44	&	13.43	&	13.44	&	13.44	\Tstrut\\
        AT	&	83.09	&	81.96	&	80.85	&	81.87	&	81.88	\\
        BA	&	15.89	&	15.44	&	15.01	&	15.34	&	15.39	\\
        BE	&	58.16	&	52.66	&	49.61	&	51.79	&	52.71	\\
        BG	&	57.30	&	57.18	&	57.02	&	57.17	&	57.18	\\
        CH	&	61.25	&	60.87	&	60.56	&	60.79	&	60.86	\\
        CZ	&	72.40	&	71.18	&	69.92	&	71.02	&	71.09	\\
        DE-LU	&	506.79	&	494.10	&	482.78	&	493.23	&	493.04	\\
        DK	&	60.47	&	60.27	&	59.53	&	60.88	&	59.93	\\
        ES	&	249.43	&	249.15	&	248.85	&	249.15	&	249.17	\\
        FI	&	98.17	&	97.88	&	97.40	&	98.02	&	97.83	\\
        FR	&	609.69	&	606.48	&	601.63	&	607.15	&	607.53	\\
        GR	&	47.32	&	47.30	&	47.29	&	47.28	&	47.29	\\
        HR	&	19.91	&	19.78	&	19.65	&	19.77	&	19.77	\\
        HU	&	43.10	&	42.96	&	42.81	&	42.93	&	42.96	\\
        IE	&	50.65	&	50.55	&	50.44	&	50.56	&	50.55	\\
        IT	&	262.94	&	260.14	&	257.57	&	259.55	&	260.04	\\
        ME	&	7.21	&	7.17	&	7.12	&	7.17	&	7.17	\\
        MK	&	7.01	&	6.99	&	6.98	&	6.99	&	6.99	\\
        NL	&	139.73	&	135.04	&	130.85	&	135.50	&	134.49	\\
        NO	&	157.06	&	157.06	&	157.07	&	157.04	&	157.06	\\
        PL	&	177.88	&	176.19	&	174.48	&	175.96	&	176.07	\\
        PT	&	53.46	&	53.41	&	53.34	&	53.42	&	53.42	\\
        RO	&	62.60	&	62.39	&	62.16	&	62.35	&	62.37	\\
        RS	&	37.53	&	37.20	&	36.91	&	37.16	&	37.17	\\
        SE	&	173.58	&	173.08	&	172.15	&	173.42	&	173.02	\\
        SI	&	19.00	&	18.78	&	18.52	&	18.77	&	18.77	\\
        SK	&	36.45	&	36.36	&	36.30	&	36.35	&	36.35	\\
        UK	&	334.18	&	332.15	&	329.37	&	332.69	&	333.04	\\
        NSEH	&	0.00	&	38.88	&	77.73	&	38.90	&	38.89	\Bstrut\\
        \hline
   \end{tabularx}}
\end{table*}

\begin{table*}[!t]
    \caption{\normalsize \textbf{Total demand} across the different scenarios (\textbf{TWh/year}).}
    \label{tabA3:demand}
    \centering
    \vspace{0.5em}
    \resizebox{!}{0.45\textwidth}{
    \begin{tabularx}{0.70\textwidth}{*1{>{\arraybackslash}X} *5{>{\centering\arraybackslash}X}}
        \hline
                  & {\bf No Hub} & {\bf 10 GW} & {\bf 20 GW} & {\bf Exchanges} & {\bf UK} \Tstrut\Bstrut\\ 
        \hline          
        AL	&	8.69	&	8.69	&	8.69	&	8.69	&	8.69	\Tstrut\\
        AT	&	86.43	&	86.84	&	87.19	&	86.91	&	86.85	\\
        BA	&	12.72	&	12.72	&	12.73	&	12.73	&	12.73	\\
        BE	&	96.01	&	96.04	&	96.31	&	95.99	&	96.02	\\
        BG	&	36.16	&	36.16	&	36.16	&	36.16	&	36.16	\\
        CH	&	65.27	&	65.42	&	65.51	&	65.44	&	65.42	\\
        CZ	&	74.46	&	74.46	&	74.46	&	74.46	&	74.46	\\
        DE-LU	&	598.05	&	599.21	&	600.48	&	599.30	&	599.13	\\
        DK	&	53.33	&	53.37	&	53.43	&	53.36	&	53.37	\\
        ES	&	274.05	&	274.16	&	274.28	&	274.17	&	274.17	\\
        FI	&	108.34	&	108.34	&	108.34	&	108.34	&	108.34	\\
        FR	&	483.57	&	483.66	&	483.76	&	483.66	&	483.66	\\
        GR	&	60.38	&	60.38	&	60.38	&	60.38	&	60.38	\\
        HR	&	18.58	&	18.58	&	18.59	&	18.58	&	18.58	\\
        HU	&	50.99	&	50.99	&	50.99	&	50.99	&	50.99	\\
        IE	&	52.98	&	52.99	&	53.00	&	52.98	&	52.91	\\
        IT	&	330.06	&	330.19	&	330.29	&	330.19	&	330.18	\\
        ME	&	4.55	&	4.55	&	4.55	&	4.55	&	4.55	\\
        MK	&	8.39	&	8.39	&	8.39	&	8.39	&	8.39	\\
        NL	&	140.52	&	140.37	&	140.71	&	140.09	&	140.37	\\
        NO	&	142.80	&	142.79	&	142.80	&	142.78	&	142.79	\\
        PL	&	182.20	&	182.20	&	182.20	&	182.20	&	182.20	\\
        PT	&	57.98	&	58.00	&	58.02	&	57.99	&	58.00	\\
        RO	&	65.09	&	65.09	&	65.09	&	65.09	&	65.09	\\
        RS	&	43.56	&	43.56	&	43.56	&	43.56	&	43.56	\\
        SE	&	153.06	&	153.06	&	153.06	&	153.06	&	153.06	\\
        SI	&	16.45	&	16.45	&	16.46	&	16.46	&	16.45	\\
        SK	&	31.42	&	31.42	&	31.42	&	31.42	&	31.42	\\
        UK	&	310.72	&	310.77	&	310.96	&	310.72	&	310.28	\\
        NSEH	&	0.00	&	0.00	&	0.00	&	0.00	&	0.00	\Bstrut\\
        \hline
   \end{tabularx}}
\end{table*}

\begin{table*}[!t]
    \caption{\normalsize \textbf{Total exports} across the different scenarios (\textbf{TWh/year}).}
    \label{tabA4:export}
    \centering
    \vspace{0.5em}
    \resizebox{!}{0.45\textwidth}{
    \begin{tabularx}{0.70\textwidth}{*1{>{\arraybackslash}X} *5{>{\centering\arraybackslash}X}}
        \hline
                  & {\bf No Hub} & {\bf 10 GW} & {\bf 20 GW} & {\bf Exchanges} & {\bf UK} \Tstrut\Bstrut\\ 
        \hline          
        AL	&	5.27	&	5.27	&	5.27	&	5.27	&	5.27	\Tstrut\\
        AT	&	13.66	&	13.11	&	12.57	&	13.11	&	13.07	\\
        BA	&	5.60	&	5.32	&	5.06	&	5.27	&	5.29	\\
        BE	&	0.07	&	0.08	&	0.07	&	0.08	&	0.08	\\
        BG	&	21.92	&	21.81	&	21.66	&	21.80	&	21.80	\\
        CH	&	13.21	&	13.01	&	12.76	&	12.96	&	13.00	\\
        CZ	&	4.32	&	3.87	&	3.39	&	3.86	&	3.82	\\
        DE-LU	&	11.31	&	9.11	&	7.48	&	9.12	&	9.21	\\
        DK	&	12.58	&	12.75	&	12.56	&	13.24	&	12.57	\\
        ES	&	17.65	&	17.52	&	17.36	&	17.51	&	17.51	\\
        FI	&	4.06	&	3.95	&	3.80	&	3.98	&	3.93	\\
        FR	&	133.14	&	130.70	&	126.86	&	131.49	&	131.80	\\
        GR	&	0.25	&	0.25	&	0.26	&	0.25	&	0.25	\\
        HR	&	4.20	&	4.12	&	4.04	&	4.11	&	4.12	\\
        HU	&	1.37	&	1.34	&	1.32	&	1.34	&	1.34	\\
        IE	&	5.90	&	5.86	&	5.81	&	5.87	&	5.93	\\
        IT	&	2.04	&	1.77	&	1.55	&	1.74	&	1.78	\\
        ME	&	3.09	&	3.06	&	3.03	&	3.06	&	3.06	\\
        MK	&	0.37	&	0.36	&	0.36	&	0.36	&	0.37	\\
        NL	&	14.14	&	12.46	&	10.95	&	13.17	&	12.13	\\
        NO	&	34.63	&	34.72	&	34.69	&	34.76	&	34.77	\\
        PL	&	5.18	&	4.60	&	4.02	&	4.58	&	4.57	\\
        PT	&	6.15	&	6.12	&	6.10	&	6.13	&	6.13	\\
        RO	&	3.26	&	3.19	&	3.11	&	3.18	&	3.18	\\
        RS	&	2.68	&	2.58	&	2.51	&	2.57	&	2.58	\\
        SE	&	24.61	&	24.40	&	23.82	&	24.73	&	24.36	\\
        SI	&	3.08	&	2.93	&	2.75	&	2.91	&	2.92	\\
        SK	&	5.39	&	5.32	&	5.26	&	5.31	&	5.31	\\
        UK	&	43.37	&	42.73	&	41.47	&	43.14	&	44.19	\\
        NSEH	&	0.00	&	38.88	&	77.73	&	38.90	&	38.89	\Bstrut\\
        \hline
   \end{tabularx}}
\end{table*}

\begin{table*}[!t]
    \caption{\normalsize \textbf{Total imports} across the different scenarios (\textbf{TWh/year}).}
    \label{tabA5:import}
    \centering
    \vspace{0.5em}
    \resizebox{!}{0.45\textwidth}{
    \begin{tabularx}{0.70\textwidth}{*1{>{\arraybackslash}X} *5{>{\centering\arraybackslash}X}}
        \hline
                  & {\bf No Hub} & {\bf 10 GW} & {\bf 20 GW} & {\bf Exchanges} & {\bf UK} \Tstrut\Bstrut\\ 
        \hline          
        AL	&	0.51	&	0.52	&	0.53	&	0.52	&	0.52	\Tstrut\\
        AT	&	7.89	&	8.51	&	9.13	&	8.60	&	8.54	\\
        BA	&	2.17	&	2.34	&	2.50	&	2.38	&	2.35	\\
        BE	&	37.53	&	43.02	&	46.25	&	43.85	&	42.96	\\
        BG	&	0.25	&	0.26	&	0.26	&	0.26	&	0.26	\\
        CH	&	12.83	&	13.00	&	13.07	&	13.04	&	13.01	\\
        CZ	&	6.30	&	7.07	&	7.86	&	7.22	&	7.11	\\
        DE-LU	&	98.65	&	109.52	&	119.64	&	110.37	&	110.67	\\
        DK	&	5.44	&	5.85	&	6.45	&	5.72	&	6.00	\\
        ES	&	33.45	&	33.59	&	33.75	&	33.59	&	33.57	\\
        FI	&	14.23	&	14.41	&	14.74	&	14.29	&	14.43	\\
        FR	&	5.32	&	6.08	&	7.13	&	6.20	&	6.14	\\
        GR	&	12.36	&	12.39	&	12.40	&	12.40	&	12.39	\\
        HR	&	2.47	&	2.52	&	2.56	&	2.52	&	2.52	\\
        HU	&	9.26	&	9.38	&	9.50	&	9.40	&	9.38	\\
        IE	&	8.03	&	8.10	&	8.17	&	8.10	&	8.09	\\
        IT	&	61.04	&	63.60	&	65.97	&	64.16	&	63.70	\\
        ME	&	0.43	&	0.45	&	0.46	&	0.45	&	0.45	\\
        MK	&	1.75	&	1.77	&	1.78	&	1.76	&	1.76	\\
        NL	&	14.92	&	17.80	&	20.82	&	17.76	&	18.01	\\
        NO	&	15.90	&	15.98	&	15.94	&	16.05	&	16.04	\\
        PL	&	9.35	&	10.46	&	11.59	&	10.66	&	10.54	\\
        PT	&	7.13	&	7.15	&	7.20	&	7.15	&	7.15	\\
        RO	&	4.02	&	4.16	&	4.31	&	4.19	&	4.17	\\
        RS	&	8.58	&	8.81	&	9.03	&	8.84	&	8.84	\\
        SE	&	4.09	&	4.37	&	4.73	&	4.37	&	4.39	\\
        SI	&	0.49	&	0.55	&	0.64	&	0.56	&	0.56	\\
        SK	&	0.18	&	0.20	&	0.20	&	0.20	&	0.20	\\
        UK	&	17.93	&	19.34	&	21.01	&	19.20	&	19.48	\\
        NSEH	&	0.00	&	0.00	&	0.00	&	0.00	&	0.00	\Bstrut\\
        \hline
   \end{tabularx}}
\end{table*}
\end{document}